\documentclass[aps,reprint, prl, superscriptaddress, longbibliography,showpacs,nobibnotes,floatfix]{revtex4-1}
\usepackage{graphicx}
\usepackage{bm, amsmath}
\usepackage{dcolumn}
\usepackage{natbib,xcolor}
\usepackage[version=4]{mhchem}
\usepackage{upgreek}
\usepackage{physics}
\begin{document}
\title{Single crystal growths and magnetic properties of hexagonal polar semimetals $\textbf{\textit{R}}$AuGe ($\textbf{\textit{R}}$ = Y, Gd$-$Tm, and Lu)}

\author{Takashi Kurumaji}
\affiliation{Department of Advanced Materials Science, University of Tokyo, Kashiwa 277-8561, Japan}
\author{Masaki Gen}
\affiliation{Department of Advanced Materials Science, University of Tokyo, Kashiwa 277-8561, Japan}
\affiliation{RIKEN Center for Emergent Matter Science (CEMS), Wako 351-0198, Japan}
\author{Shunsuke Kitou}
\affiliation{RIKEN Center for Emergent Matter Science (CEMS), Wako 351-0198, Japan}
\author{Kazuhiko Ikeuchi}
\affiliation{Neutron Science and Technology Center, Comprehensive Research Organization for Science and Society (CROSS), Tokai, Ibaraki, 319-1106, Japan}
\author{Mitsutaka Nakamura}
\affiliation{J-PARC Center, Japan Atomic Energy Agency, Tokai, Ibaraki 319-1195, Japan}
\author{Akihiko Ikeda}
\affiliation{Department of Engineering Science, University of Electro-Communications, Chofu, Tokyo 182-8585, Japan}
\author{Taka-hisa Arima}
\affiliation{Department of Advanced Materials Science, University of Tokyo, Kashiwa 277-8561, Japan}
\affiliation{RIKEN Center for Emergent Matter Science (CEMS), Wako 351-0198, Japan}
\date{\today}
\begin{abstract}

We study structural and magnetic properties of rare-earth based semimetals $\textit{R}$AuGe ($\textit{R}$ = Y, Gd$-$Tm, and Lu) using flux-grown single crystals.
These compounds belong to the noncentrosymmetric polar space group $P6_{3}mc$.
We confirm the systematic structural evolution at room temperature as a function of ionic radius of rare earths to clarify the isopointal crossover between two polar structures: three-dimensional LiGaGe-type and quasi-two-dimensional NdPtSb-type.
Magnetism shows a characteristic anisotropy in reasonable agreement with the crystal electric field (CEF) theory; the easy-plane-type anisotropy for $\textit{R}$ = Tb and Dy turns into the Ising-type anisotropy for $\textit{R}$ = Er and Tm.
We evaluate the CEF parameters based on the Stevens operators to reasonably reproduce the temperature dependence of magnetic susceptibilities and specific heat for $\textit{R}$AuGe ($\textit{R}=$ Tb$-$Tm).
The estimated energy scale of the Ising gap ($\sim 11$ meV) in TmAuGe is consistent with an excitation observed in an inelastic neutron scattering experiment.
%The compounds with $\textit{R}$ = Gd$-$Er are antiferromagnetic below the transition temperature ranging from 15 K for Gd to 3.2 K for Er, while TmAuGe is ferromagnetic below $T_{\text{C}}=4.1$ K.
%GdAuGe and HoAuGe undergo successive phase transitions as sweeping the temperature.
%The temperature dependence of magnetic susceptibility for $\textit{R}$ = Er and Dy are hysteretic suggesting glassy magnetism and multidomain nature owing to the breaking of $C_6$ rotational symmetry of crystal, respectively.
These findings suggest an opportunity for interplay between conduction electrons and nontrivial spin structures in the family of magnetic polar semimetals $\textit{R}$AuGe.
\end{abstract}

\maketitle

\section{Introduction}
Polar metals are defined as the conductive materials with symmetry breaking allowing electric polarization, which provide a possible route to ferroelectric metals \cite{shi2013ferroelectric,benedek2016ferroelectric} and to hyperferroelectricity \cite{garrity2014hyperferroelectrics,zhou2020review}.
These materials are potentially applied to electric-field-switchable spintronic devices \cite{narayan2015class,fei2018ferroelectric}, and a metallic counterpart of multiferroics \cite{cao2017dirac,gao2018dirac,gao2021noncentrosymmetric} with topological band crossings induced by magnetic correlation \cite{du2015dirac,mondal2019broken} and/or spin textures \cite{gaudet2021weyl,srivastava2020observation,zhang2022room,zhang2022room2}.

Rare-earth-based semimetals $\textit{R}$AuGe ($\textit{R}=$ Sc, Y, La$-$Nd, Sm, Gd$-$Tm, and Lu) belong to the equiatomic ternary $\textit{RTX}$ ($\textit{T}=$ transition metals, $X=$ p-block elements) intermetallic phases \cite{gupta2015review} of the hexagonal polar space group $P6_{3}mc$ (No. 186) \cite{rossi1992ternary,pottgen1996crystal}.
Their crystal structure is shown in Fig. 1(a), where the positions of Au and Ge are shifted upward and downward, respectively, along the $c$ axis from the aristotype ZrBeSi structure (space group $P6_3/mmc$, No. 194) to break the inversion symmetry, which is in common with a prototypical polar metal considered in theories \cite{bennett2012hexagonal,garrity2014hyperferroelectrics,gibson2015three,du2015dirac,di2016intertwined,cao2017dirac,gao2018dirac,mondal2019broken,zhang2019influences,gao2021noncentrosymmetric}.

CeAuGe and TmAuGe are known to show ferromagnetism below $T_\text{C}=10$ K \cite{pottgen1996ferromagnetic,gibson2000magnetic,gibson1996susceptibility} and $T_\text{C}=4$ K \cite{kaczorowski2014magnetic}, respectively.
The compounds for other $\textit{R}$ (Nd, and Gd$-$Er) are antiferromagnetic with a transition temperature $T_{\text{N}}$ ranging between 6 K and 17 K \cite{gibson1996susceptibility,gibson1998investigation,penc1999magnetic,baran2000magnetic,baran2001neutron,gibson2001crystal,bashir2014antiferromagnetic}.
A previous neutron study detected a commensurate modulation vector $(\frac{1}{2},0,0)$ in the ground state for $\textit{R}=$ Tb$-$Er \cite{gibson1998investigation,baran2000magnetic,baran2001neutron,gibson2001crystal}.
Incommensurate modulations were also observed to coexist at low temperatures or just below $T_{\text{N}}$, suggesting inherent magnetic frustration in triangular-lattice layers of $\textit{R}$.
The density of state has a pseudo-gap nature near the Fermi level due to the overlap of electron and hole bands, which is enabled by nominal valence configuration $\textit{R}^{3+}$Au$^{+}$Ge$^{4-}$ \cite{pottgen1996crystal,schnelle1997crystal,wang2020chemical} and contributes to semimetallic transport properties of nonmangetic compounds ($\textit{R}=$ Sc, Y, La, and Lu) \cite{schnelle1997crystal,gibson1996susceptibility}.
Resistivities of magnetic compounds, $\textit{R}$AuGe ($\textit{R}=$ Ce, Nd, Gd, Ho, and Tm) \cite{pottgen1996ferromagnetic,gibson2001crystal,kaczorowski2014magnetic,bashir2014antiferromagnetic}, are known to show nonmonotonic temperature dependences indicating a coupling with magnetic orders. 
These features suggest an opportunity to encounter unconventional properties of polar semimetals intertwinned with frustrated magnetism in magnetic $\textit{R}$AuGe.
The above-mentioned physical properties were identified using polycrystals, and single crystal growths are only known for CeAuGe using Czochlarski and floating-zone methods \cite{gibson1998investigation,prokofiev2012crystal}, and for epitaxial thin-film form of LaAuGe \cite{du2019high}.

In this study, we report the single-crystal growths and physical properties of $\textit{R}$AuGe ($\textit{R}=$ Y, Gd$-$Tm, and Lu).
We analyzed the crystal structure using x-ray radiation to confirm the systematic evolution of polar structure as a function of $\textit{R}$-ionic radius.
We measured specific heat, magnetization, and thermal expansion to reveal the anisotropic magnetism significantly affected by the crystal electric field (CEF) as well as magnetic transitions at low temperatures.
We estimate the CEF potential parameters to reproduce the magnetic susceptibility and specific heat and obtain a systematic $\textit{R}$ dependence reasonably agreed with the Stevens theory \cite{Stevens}.
The estimated Ising gap in TmAuGe is 11 meV, which is consistent with the excitation at 13 meV observed in the inelastic neutron scattering (INS) experiment.
These results provide basic understanding on the structural and magnetic properties of the family of polar semimetals $\textit{R}$AuGe.

\begin{figure}[t]
	\includegraphics[width = \columnwidth]{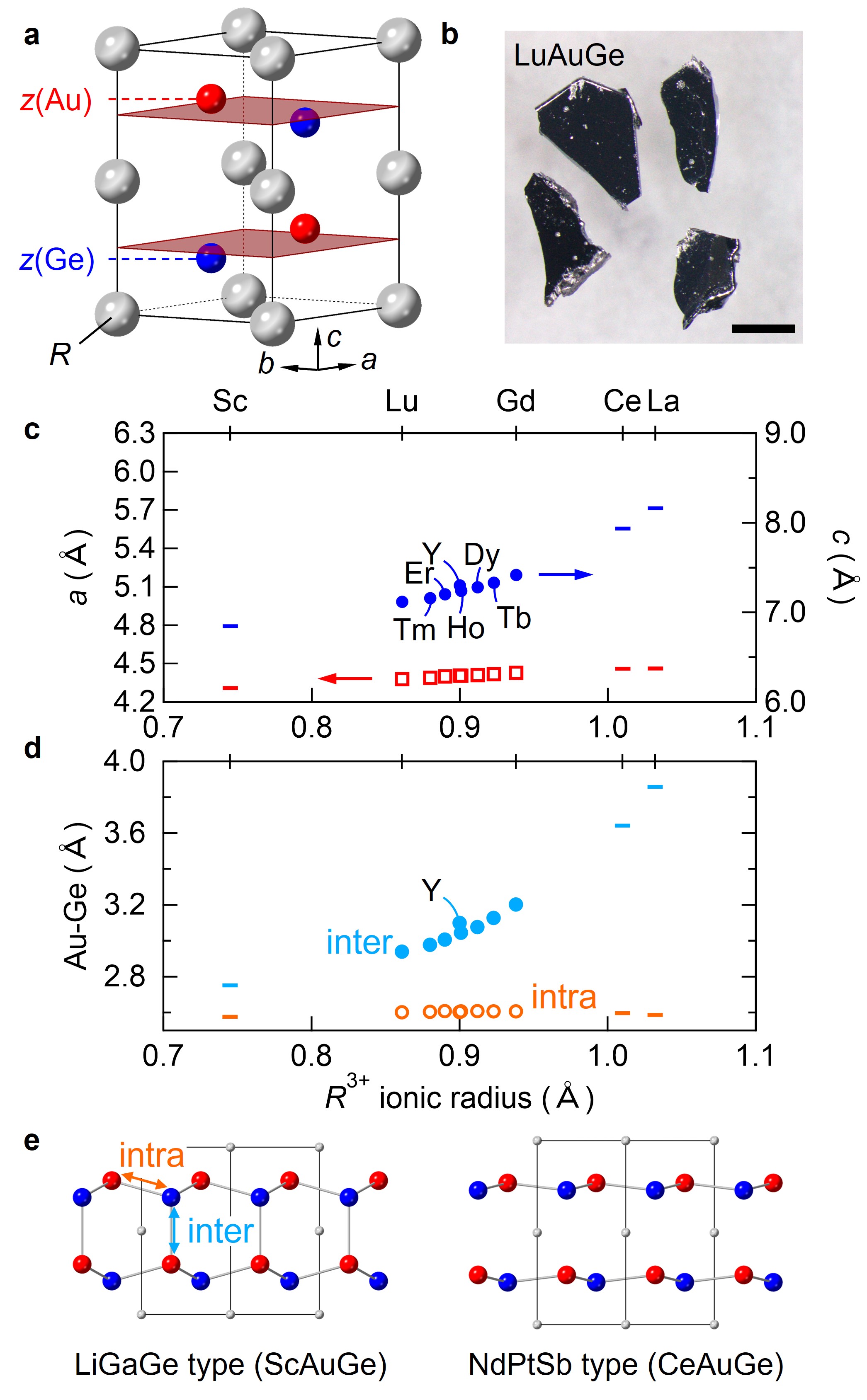}
	\caption{\label{fig1}(a) Crystal structure of $\textit{R}$AuGe.
	$z$(Au) and $z$(Ge) are fractional coordinates along the $c$ axis of Au and Ge atoms at $2b$ sites $(\frac{1}{3},\frac{2}{3},z)$, which are shifted upward from 0.75 and downward from 0.25, respectively, producing the polarity in $\textit{R}$AuGe.
	(b) Photograph of single crystals of LuAuGe.
	Black bar is 1 mm.
	(c)-(d) Lattice constants, $a$ and $c$, and inter and intralayer Au-Ge bond lengths in a unit cell at room temperature as a function of the $\textit{R}^{3+}$ ionic radius for six-coordination.
	The values for LaAuGe are excerpts from Ref. \cite{schnelle1997crystal}, and those for CeAuGe and ScAuGe are from Ref. \cite{pottgen1996crystal}.
	(e) Side views of two isopointal \cite{pottgen1996crystal} structures, LiGaGe and NdPtSb types, with distinct interlayer Au-Ge bonding (cyan arrow).
}
\end{figure}

\begin{figure}[t]
	\includegraphics[width =  0.7\columnwidth]{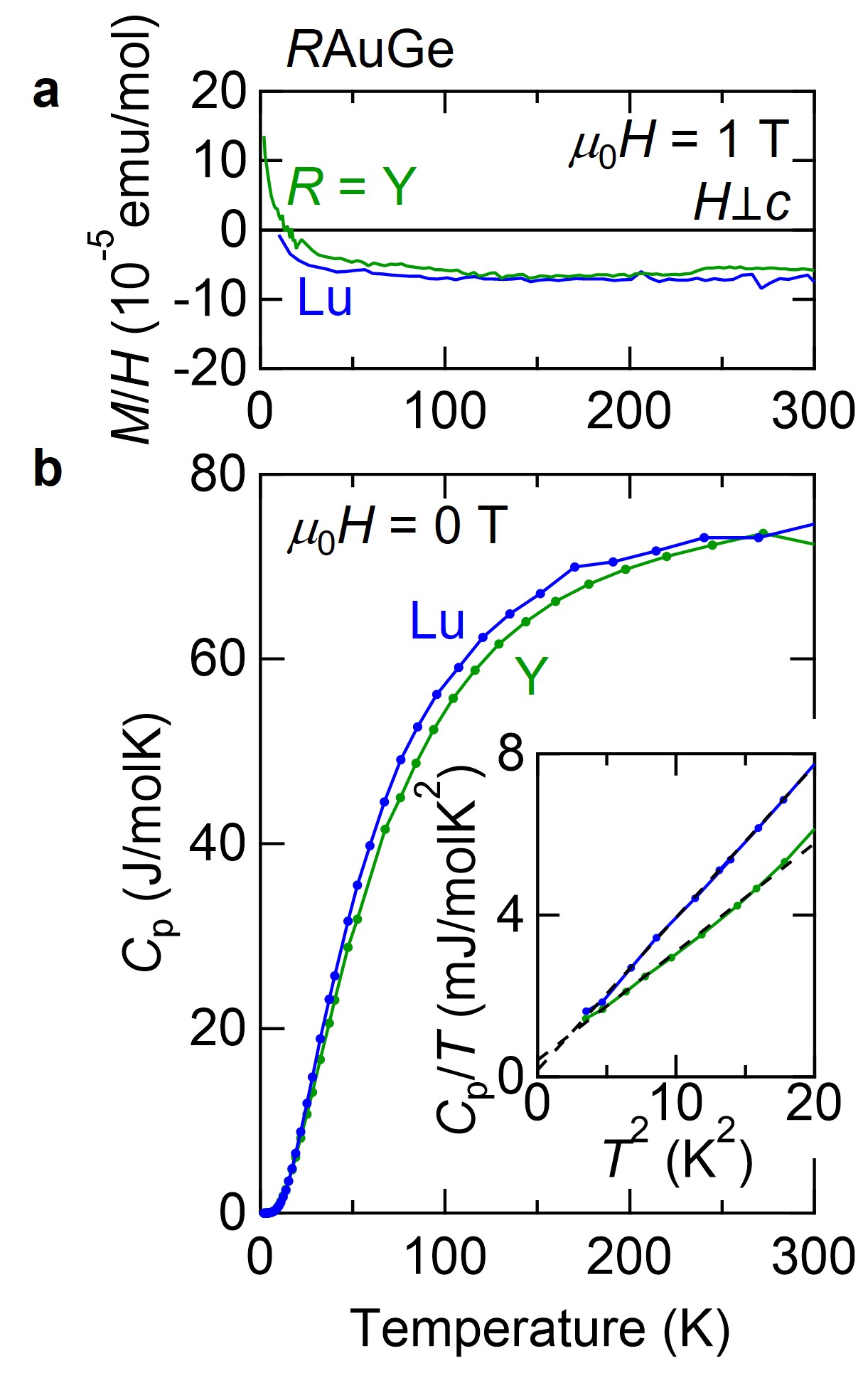}
	\caption{\label{fig2}(a)-(b) Temperature dependence of magnetic susceptibility ($M/H$) in $H\perp c$ and the specific heat ($C_{\text{p}}$) of $\textit{R}$AuGe ($\textit{R}$ = Y and Lu) at zero field.
	The inset of (b) plots $C_{\text{p}}/T$ vs. $T^2$.
	The dashed lines are linear fit.
	}
\end{figure}

\section{Experimental Methods}
Single crystals of $\textit{R}$AuGe ($\textit{R}$ = Y, Gd$-$Tm, and Lu) were grown using the Au-Ge self-flux method.
The molar ratio of $\textit{R}$:Au:Ge was 1:2:2 for $\textit{R}$ = Y, Tb$-$Tm, and Lu and 1:3:3 for $\textit{R}$ = Gd.
Starting materials were $\textit{R}$ ingot (99.9\%), Au wire (99.5\%), and Ge chunks (99.999\%).
They were loaded into an alumina crucible, which was sealed in an evacuated quartz tube.
The tube was heated to 1100 $^{\circ}$C and kept at this temperature for 24 hours before being slowly cooled to 800 $^{\circ}$C in 200 hours.
After being kept at 800 $^{\circ}$C for 4 days, the tube was taken out from the furnace and centrifuged to remove excess flux.
The obtained crystals were of platelet shape with typical dimensions 3$\times$3$\times$0.5 mm$^{3}$ as shown in Fig. 1(b). 
The atomic composition and the phase were checked by energy dispersive x-ray spectroscopy (EDS, JEOL model JSM-6010LA) and a powder x-ray diffraction (Rigaku SmartLab diffractometer) using Cu $K_{\alpha}$ radiation, respectively.
Single-crystal X-ray diffraction data for $\textit{R}$AuGe were collected on Rigaku AFC-8 diffractometer equipped with a HyPix-6000 detector using Mo $K_{\alpha}$ radiation at room temperature.
The intensities of Bragg reflections were collected by CrysAlisPro program \cite{crysalispro2014agilent}.
The crystal structures were solved by SUPERFLIP \cite{palatinus2007superflip} and refined by Jana2006 \cite{petvrivcek2014crystallographic}.
The details of the structural analysis results are shown in supplementary materials.
Attempts to grow the $P6_3mc$ phase for YbAuGe \cite{rossi1992ternary} were unsuccessful and resulted in the orthorhombic phases, consistent with $\alpha$-, $\beta$- (space group $Pnma$, No. 62) or $\gamma$-YbAuGe (space group $Imma$, No. 74) as reported in Ref. \cite{merlo1998phases}.

Specific heat $C_{\text{p}}$ was measured using a commercial system (heat capacity option of a Quantum Design PPMS).
Magnetization was measured using a superconducting quantum interference device magnetometer (Quantum Design MPMS-XL).
Thermal expansion was measured by the fiber-Bragg-grating (FBG) technique using an optical sensing instrument (Hyperion si155, LUNA) in a cryostat equipped with a superconducting magnet (Oxford Spectromag).
Optical fibers were glued using epoxy (Stycast1266) on a (001) surface of as-grown crystals to measure the elongation or compression along the $ab$ plane.

\begin{figure}[t]
	\includegraphics[width =  \columnwidth]{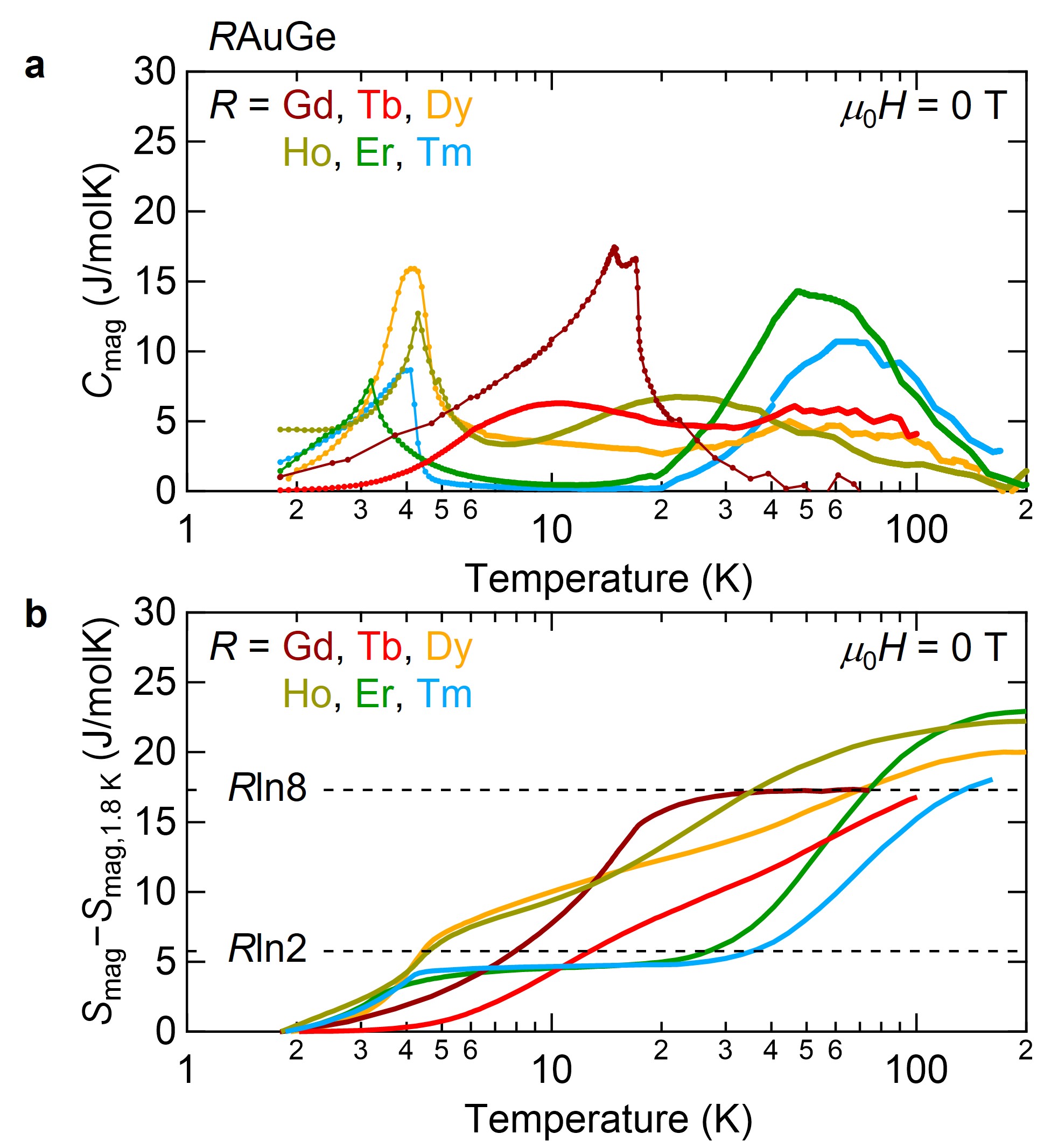}
	\caption{\label{fig3}(a)-(b) Temperature dependence of magnetic part of specific heat ($C_{\text{mag}}$) and that of entropy ($S_{\text{mag}}$) of $\textit{R}$AuGe ($\textit{R}$ = Gd$-$Tm) in zero field.
	LuAuGe data was used to subtract the phonon background. 
}
\end{figure}

\begin{figure}[t]
	\includegraphics[width =  \columnwidth]{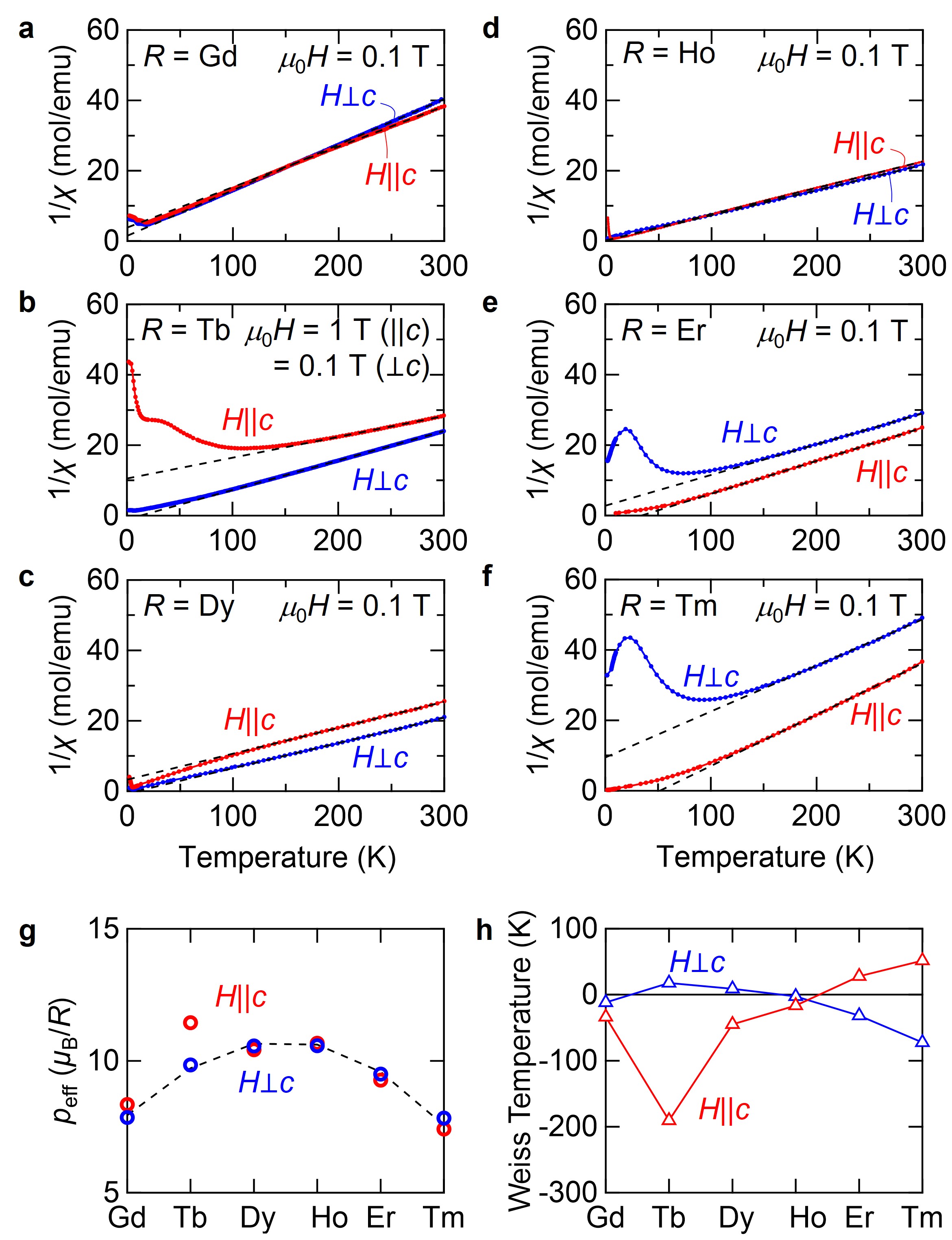}
	\caption{\label{fig4}(a)-(f) Temperature dependence of the inverse magnetic susceptibility ($H/M$) of $\textit{R}$AuGe ($\textit{R}$ = Gd$-$Tm) for $H\parallel c$ (red) and $H\perp c$ (blue).
	Dashed lines are Curie-Weiss fit to the data in $T=$ 150-300 K.
	(g)-(h) $\textit{R}$ dependence of (g) effective magnetic moment ($p_{\text{eff}}$), and (h) Weiss temperature.
	Dashed line in (a) is free ions values.
}
\end{figure}

INS experiments were performed on a time-of-flight (TOF) spectrometer (4SEASONS) at the Materials and Life Science Experimental Facility (MLF) of J-PARC in Japan \cite{4seasons}.
Polycrystalline \ce{TmAuGe} sample of 5 g was prepared by arc melting of elements.
A boule of \ce{TmAuGe} was annealed at 500 $^{\circ}$C for 21 days in an evacuated quartz tube, and sliced into pieces with thickness of 3 mm to reduce the absorption of neutrons.
The chopper frequency was set to 200 Hz, and the incident energy of $E_{\text{i}} = 41$ meV was used to observe low-energy excitations.
The sample was set in a GM refrigerator to control the temperature down to 5 K.
The aquired data were analyzed with the UTSUSEMI software package \cite{UTSUSEMI}.

\section{Results}
The crystal structures of $\textit{R}$AuGe are systematically dependent on the ionic radius of $\textit{R}$.
Single-crystal x-ray diffraction confirms the polar $P6_3mc$ structure for all $\textit{R}$, consistent with the previous reports \cite{rossi1992ternary,pottgen1996crystal,schnelle1997crystal}.
Figures 1(c) and (d) display the $\textit{R}^{3+}$ ionic radius dependence of the lattice constants and inter- and intralayer Au-Ge bond lengths.
Larger $\textit{R}$ ions lead to the expansion of the unit-cell volume especially resulting in the elongation of the $c$ axis.
This tendency correlates with the weakening of puckered distortion of Au-Ge honeycomb layer, where the out-of-plane (interlayer) Au-Ge bond becomes longer (see Fig. 1(d)).
ScAuGe has the smallest $\textit{R}$ radius (Sc$^{3+}$(VI):0.745 \AA) with the interlayer Au-Ge bond of 275.2 pm, while it elongates to 385.8 pm in LaAuGe with the largest $\textit{R}$-ion radius (La$^{3+}$(VI):1.032 \AA) \cite{shannon1976revised}.
The materials for $\textit{R}=$ Y, Gd$-$Tm, and Lu smoothly interpolate the two extremes to verify the isopointal crossover of the crystal-structure type between NdPtSb type and LiGaGe type (Fig. 1(e)), both of which belong to the space group $P6_3mc$ \cite{pottgen1996crystal,pottgen1996ferromagnetic,hoffmann2001alb2}.
In other words, as the ionic $\textit{R}^{3+}$ becomes larger, Au-Ge networks evolve from three-dimensional ZnO-type to staggered stacking of two-dimensional puckered binary honeycomb layers, asymptotically towards hBN-type network in the centrosymmetric ZrBeSi-type structure (space group $P6_3/mmc$, No. 194).

Next, we look through the physical properties of nonmagnetic YAuGe and LuAuGe.
Magnetic susceptibility (Fig. 2(a)) shows diamagnetic nature for both compounds at high temperatures, which is consistent with the previous report \cite{schnelle1997crystal}.
The low-temperature rise is ascribable to magnetic impurities.
Specific heat $C_{\text{p}}$ (Fig. 2(b)) is dominated by lattice contribution while $C_{\text{p}}$ for LuAuGe is larger than that of YAuGe in the entire temperature range because of the lower Debye temperature \cite{schnelle1997crystal}.
Small electronic contribution, $\gamma T$, to $C_{\text{p}}$ is deduced by $C_{\text{p}}/T$-vs.-$T^2$ plot (see inset of Fig. 2(b)).
Densities of states at Fermi energy per a formula unit for YAuGe and LuAuGe are 0.175 eV$^{-1}$ ($\gamma =$ 0.4 mJ/mol K$^2$) and 0.085 eV$^{-1}$ ($\gamma =$ 0.2 mJ/mol K$^2$), respectively, the order of magnitude of which agrees with the previous study \cite{schnelle1997crystal}.

\begin{figure*}[t]
	\includegraphics[width =  \textwidth]{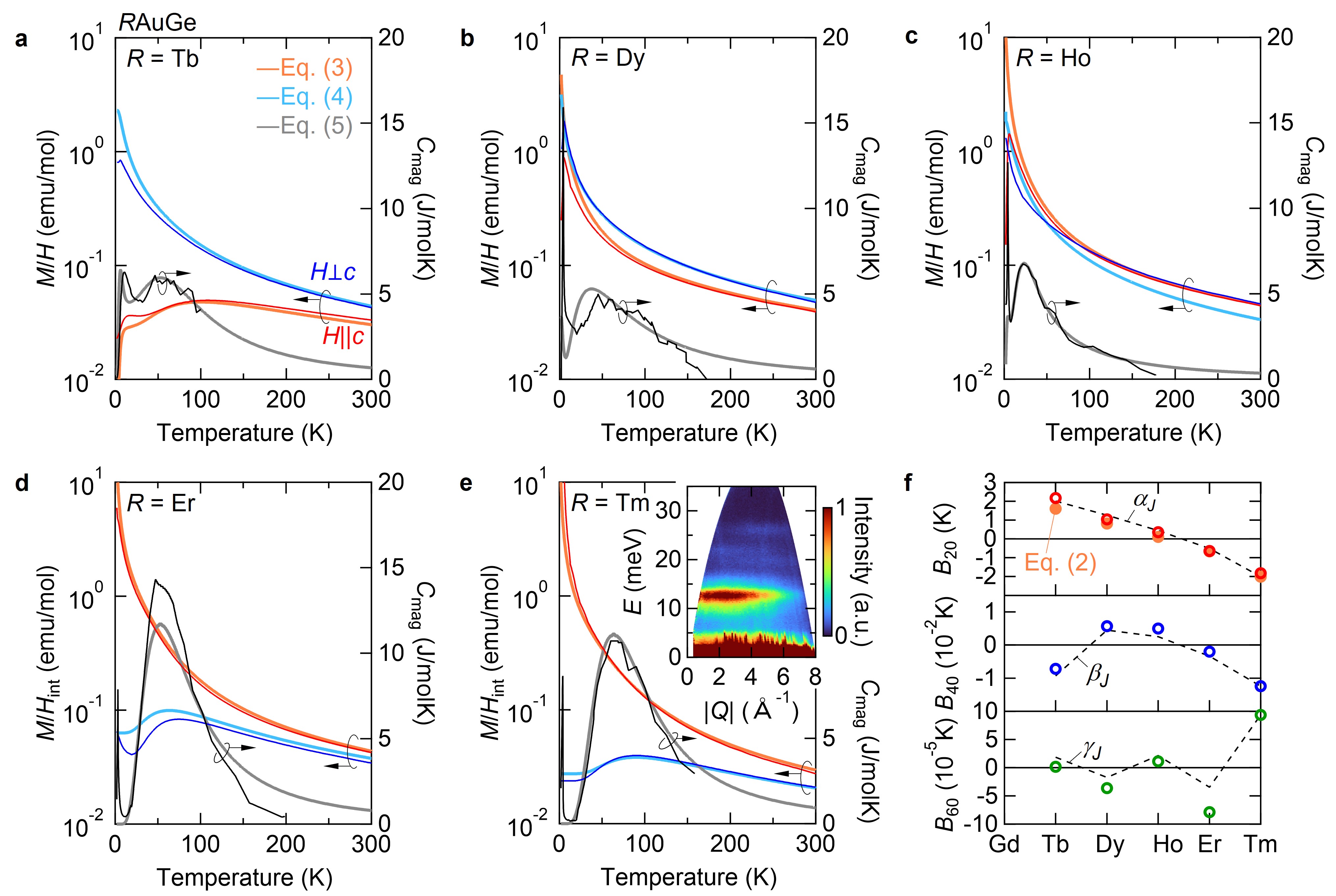}
	\caption{\label{fig5} (a)-(e) Temperature dependence of $M/H$ for $H\parallel c$ (red) and $H\perp c$ (blue), and $C_{\text{mag}}$ (black) of $\textit{R}$AuGe ($\textit{R}=$ Tb$-$Tm).
	Magnetic susceptibility of ErAuGe and TmAuGe for $H\parallel c$ (red curves in (d) and (e)) is corrected by the demagnetization factor, $N_{\text{d}}$, where $M$ is divided by the internal magnetic field, $H_{\text{int}} (=H-N_{\text{d}}M$).
	Cyan, orange, and gray curves are obtained from Eqs. (3)-(5), respectively (see text) with optimized parameters, $B_{20}$, $B_{40}$, and $B_{60}$ (see supplementary materials).
	Inset of (e) is a color map of intensity vs. $|Q|$ of the polycrystalline INS profile of TmAuGe with the incident neutron energy $E_{\text{i}} =$ 41 meV at $T=$ 5 K.
	(f) $\textit{R}$ dependence of $B_{20}$, $B_{40}$, and $B_{60}$ (open circles).
	Dashed lines are Stevens factors, $\alpha _J$, $\beta _J$, and $\gamma _J$ \cite{Stevens}, normalized by the values for $\textit{R}=$ Tm.
	Closed orange circles in (f) is obtained from the anisotropy of Curie-Weiss curves using Eq. (2).
}
\end{figure*}

Using $C_{\text{p}}$ of LuAuGe as the reference for the lattice contribution, we obtain the magnetic specific heat ($C_{\text{mag}}$) for $\textit{R}=$ Gd$-$Tm (Fig. 3(a)).
The double peak structure at $T=15$ and 17 K for $\textit{R}=$ Gd reproduces the previous measurement \cite{gibson1996susceptibility}. 
The peaks below 10 K for $\textit{R}=$ Dy$-$Tm represent magnetic transitions at low temperatures while a Schottky peak around 50-60 K for $\textit{R}=$ Er and Tm, for example, suggests the significant contribution of crystal electric field (CEF) excitations.
To gain insight on the CEF energy level, we integrate $C_{\text{mag}}/T$ with respect to $T$ to estimate the magnetic entropy $S_{\text{mag}}$ (Fig. 3(b)). 
For $\textit{R}=$ Gd, the most of $S_{\text{mag}}$ for free-spin value, $R\ln8$, is released below the transition temperatures owing to the Heisenberg nature of Gd spins.
The release of $S_{\text{mag}}$ between 4 and 20 K for ErAuGe and TmAuGe only approaches $R\ln2$, suggesting the contribution of the lowest-lying (quasi-)doublet well isolated from excited states.
This feature reflects on the Ising-type anisotropy of their magnetic moments.
For $\textit{R}=$ Tb$-$Ho, in contrast, the $T$-dependence of $S_{\text{mag}}$ does not show any saturation behavior below 100 K, suggesting overlapped contributions of CEF excitations.

\begin{table*}
\centering
\caption{\label{table1}Structural and magnetic parameters of $\textit{R}$AuGe ($\textit{R}$ = Y, Gd$-$Tm, and Lu).
$a$, $c$: lattice constants, and $z_{\text{Au}}$, $z_{\text{Ge}}$: fractional coordinate for Au and Ge atoms along the $c$ axis (see supplementary materials).
$T_{\text{C}}$: ferromagnetic transition temperature. $T_{\text{N}}$: antiferromagnetic transition temperature. $p_{\text{eff}}$: effective magnetic moment, and $\mathit{\Theta} _{\text{W}}$: Weiss temperature for $H\parallel c$ and $H\perp c$.}
\begin{tabular}{l*{8}{c}}
\hline
\hline
$\textit{R}$ & Y & Gd & Tb & Dy & Ho & Er & Tm & Lu  \\
\hline
\multicolumn{9}{c}{Structural parameters}\\
\hline
$a$ (\AA) & 4.4061(2) & 4.42799(1) & 4.4177(2) & 4.4105(3) & 4.4045(2) & 4.3996(2) & 4.3879(3) & 4.3772(3) \\
$c$ (\AA) & 7.3011(4) & 7.4176(2) & 7.3300(3) & 7.2801(4) & 7.2402(4) & 7.2018(3) & 7.1594(5) & 7.1175(3) \\
$z_{\text{Au}}$ & 0.78435(5) & 0.77968(6) & 0.78308(7) & 0.78614(11) & 0.78703(6) & 0.78931(6) & 0.79083(7) & 0.79415(9) \\
$z_{\text{Ge}}$ & 0.2088(2) & 0.2114(3) & 0.2097(3) & 0.2088(6) & 0.2075(3) & 0.2069(3) & 0.2068(3) & 0.2071(5) \\
\hline
\multicolumn{9}{c}{Magnetic properties}\\
\hline
$T_{\text{C}}$ (K) & - & - & - & - & - & - & 4.1 & - \\
$T_{\text{N}}$ (K) & - & 17.0 & 6.0 & 4.4 & 4.9 & 3.2 & - & - \\
$p_{\text{eff}\perp c}$ & - & 7.9 & 9.8 & 10.6 & 10.6 & 9.5 & 7.8 & - \\
$\mathit{\Theta}_{\text{W}\perp c}$ (K) & - & -11.6 & 17.6 & 9.0 & -2.6 & -31.8 & -72.3 & - \\
$p_{\text{eff}\parallel c}$ & - & 8.3 & 11.4 & 10.4 & 10.7 & 9.5 & 7.4 & - \\
$\mathit{\Theta}_{\text{W}\parallel c}$ (K) & - & -33.8 & -191 & -45.0 & -16.6 & 33.0 & 51.6 & - \\
\hline
\hline
\end{tabular}
\end{table*}

The systematic $\textit{R}$-dependence of magnetic anisotropy can be seen in the inverse magnetic susceptibility ($H/M$) summarized in Figs. 4(a)-(f).
Magnetic properties for $\textit{R}=$ Gd and Ho are roughly isotropic showing similar in- and out-of-plane $M/H$. 
For $\textit{R}=$ Tb and Dy, easy-plane-type anisotropy is observed, while ErAuGe and TmAuGe host easy-axis-type anisotropy.
All the compounds show a Curie-Weiss behavior at high temperatures, where the Curie constants and the Weiss temperatures ($\mathit{\Theta} _{\text{W}}$) are obtained by a linear fit of the $H/M$-$T$ curve between 150 and 300 K.
The effective moment $p_{\text{eff}}$ is evaluated from the Curie constant.
The $\textit{R}$ dependences of $p_{\text{eff}}$ and $\mathit{\Theta} _{\text{W}}$ are summarized in Figs. 4(g) and (h).
Except for $\textit{R}=$ Tb, $p_{\text{eff}}$ agrees well with the free-ion value (dashed curve in Fig. 4(g)), suggesting that the energy scale of CEF is lower than the thermal energy at room temperature.
$\mathit{\Theta} _{\text{W}}$ for $H\parallel c$ is larger than that for $H\perp c$ in the case of $\textit{R} =$ Er and Tm.
The systematic anisotropy change is associated with the CEF parameters \cite{Stevens}.

We discuss the effect of CEF to clarify the $\textit{R}$ dependence of magnetic anisotropy in $\textit{R}$AuGe. 
The 4f electrons on $\textit{R}$ sites in $\textit{R}$AuGe are affected by the CEF potential ($V_{\text{CEF}}$) for the site symmetry of $C_{3v}$.
$V_{\text{CEF}}$ can be expressed by the Stevens operators ($\hat{O}_n^m$) as
\begin{align}
\label{VCEF}
V_{\text{CEF}}=B_{20}\hat{O}_2^0+B_{40}\hat{O}_4^0+B_{60}\hat{O}_6^0\notag\\
+B_{43}\hat{O}_4^3+B_{63}\hat{O}_6^3+B_{66}\hat{O}_6^6. \\
\notag
\end{align} 
Here, we assume $B_{43}=B_{63}=B_{66}=0$ for simplicity of calculations, which is reasonable when the axial anisotropy of magnetic properties mainly arises from $B_{20}$, $B_{40}$, and $B_{60}$.
In such a case the $\ket{J_z = n}$ state is also an eigenstate of $V_{\text{CEF}}$.
In the high temperature limit, the $\hat{O}_2^0 = [3\hat{J}_z^2-J(J+1)]$ term dominates the anisotropy of magnetic susceptibilities \cite{wang1971crystal,boutron1973exact}.
The difference between the inverse of magnetic susceptibility for $H\parallel c$ and $H\perp c$, $\chi _{\parallel c} ^{-1}$ and $\chi _{\perp c} ^{-1}$, respectively, is approximated to be a constant as
\begin{equation}\label{Eq4}
\chi_{\parallel c}^{-1}-\chi_{\perp c}^{-1}=\frac{3}{2}\frac{(2J-1)(2J+3)}{5C_{\text{ave}}}B_{20},
\end{equation}
where $J$ is the total angular momentum $J=L\pm S$ and $C_{\text{ave}}$ is the Curie constant obtained by their average for $H\parallel c$ and $H\perp c$.
Assuming the identical radial distribution of 4f electron wave function for different $\textit{R}$, $B_{20}$ is predicted to be scaled with a Stevens factor $\alpha _J$ \cite{Stevens}, which smoothly evolves from $\textit{R}=$ Tb to Tm and changes the sign through Ho to Er.
This feature of $B_{20}$ agrees well with the $\textit{R}$-dependence of magnetic susceptibility capturing the origin of the anisotropy switch in $\textit{R}$AuGe, which will be discussed more quantitatively later.

To evaluate the systematic evolution of the CEF of $\textit{R}^{3+}$ ions, we simulate the temperature dependence of $\chi_{\parallel c}$, $\chi_{\perp c}$, and $C_{\text{mag}}$ on the basis of the formulae (see e.g. Ref. [\onlinecite{van2007magnetic}]):

\begin{multline}
\chi_{\parallel c}=\frac{N_{\text{A}}(g_J\mu _{\text{B}})^2}{Z}\notag\\
\qquad\qquad\times \frac{1}{k_{\text{B}}T}\sum_{n=-J}^{J}J_z^2 e^{-E_n/k_{\text{B}}T}, \qquad\qquad\qquad\qquad   (3)\\
\end{multline}

\begin{multline}
\chi_{\perp c}=\frac{2N_{\text{A}}(g_J\mu _{\text{B}})^2}{Z}\notag \\
\times  \sum_{n=-J}^{J-1}\abs{\mel{n+1}{J_x}{n}}^2\frac{e^{-E_n/k_{\text{B}}T}-e^{-E_{n+1}/k_{\text{B}}T}}{E_{n+1}-E_n}, \qquad  (4)\\
\notag
\end{multline}

\begin{multline}
\label{Cmag}
C_{\text{mag}}=\frac{N_{\text{A}}}{k_{\text{B}}T^2}\left[ \left(\frac{1}{Z}\sum_{n} E_n^2e^{-E_n/k_{\text{B}}T} \right)\right. \notag\\
\left.\qquad\qquad\qquad - \left( \frac{1}{Z}\sum_{n} E_n e^{-E_n/k_{\text{B}}T}\right)^2 \right],\qquad\qquad      (5)\\
\notag
\end{multline}
where $N_{\text{A}}$ is Avogadro constant, $k_{\text{B}}$ is Boltzmann constant, $\mu_{\text{B}}$ is Bohr magneton, $g_{J}$ is Lande's $g$ factor, $E_{n}$ is the energy for a state, $\ket{J_z = n}$.
$Z=\sum _{n}e^{-E_n/k_{\text{B}}T}$ is the partition function.
We summarize the results of simulation of magnetic susceptibilities and magnetic specific heat in Figs. 5(a)-(e), and the deduced Stevens parameters in Fig. 5(f).
We note that all the Stevens parameters systematically scale with the Stevens factors, $\alpha_J$, $\beta_J$, and $\gamma_J$, confirming the validity of the simulations.
We also show the $B_{20}$ values estimated from the formula of high-$T$ limit Eq. (2) (closed orange circle) in Fig. 5(f), agreeing well with the all-data analysis.

\begin{figure*}[t]
	\includegraphics[width =  \textwidth]{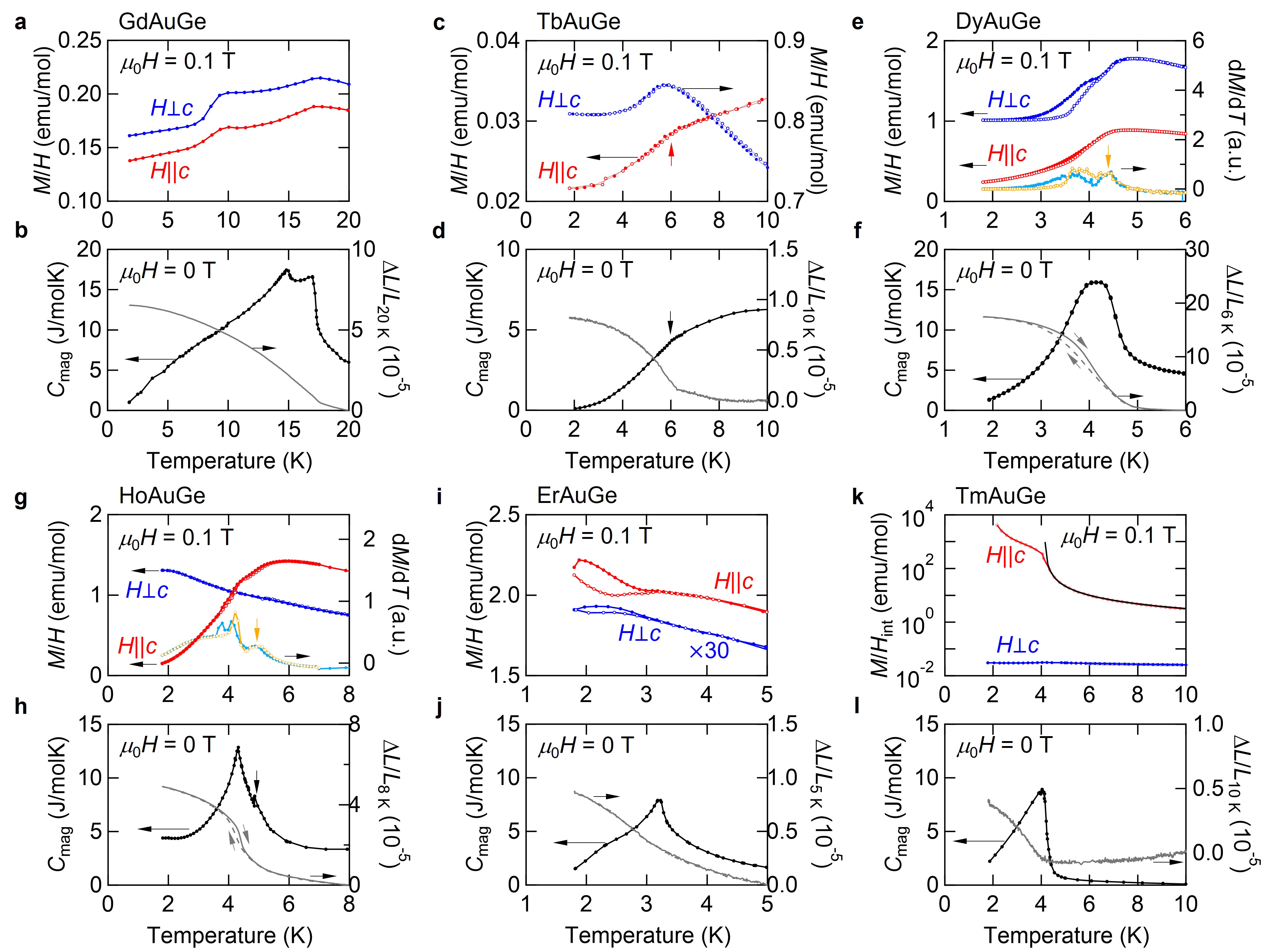}
	\caption{\label{fig6} (a)-(l) Temperature dependence of magnetic and thermodynamic properties of $\textit{R}$AuGe ($\textit{R}=$ Gd$-$Tm) at low temperatures.
	Red and blue curves represent $M/H$ for $H\parallel c$ and $H\perp c$ at $\mu _0H=$ 0.1 T, respectively.
	Black and gray curves represent $C_{\text{mag}}$ and in-plane $\Delta L/L_0$ in zero field.
	In (e) and (g), temperature derivative of $M$ ($\text{d}M/\text{d}T$) for $H\perp c$ and $H\parallel c$, respectively, are shown.
	Closed (open) symbols and solid (dashed) lines are measured in temperature-decrease (increase) processes.
	In (k), demagnetization field is corrected by using the internal magnetic field, $H_{\text{int}}=H_{\text{ext}}-N_{\text{d}}M$.
	Black curve in (k) is the Curie-Weiss fit ($\chi_{\parallel c}=\frac{C_{\text{low}}}{T-\mathit{\Theta}_{\parallel c\text{,low}}}$) for 4 K $< T <$ 30 K.
}
\end{figure*}

To double check the CEF gap in TmAuGe, we perform INS and observe the energy transfer as shown in the inset of Fig. 5(e).
Crystal field around a Tm$^{3+}$ ion in $C_{3v}$ symmetry splits the $J = 6$ manifold of Tm$^{3+}$ 4f$^{12}$ levels into five singlets and four doublets.
The lowest quasi-doublet is mainly composed of $J_z = \pm 6$, and the CEF splitting between the quasi-doublet and $J_z=\pm 5$ doublet provides Ising nature of the magnetic moment.
At 5 K, dispersionless excitations are observed at around $13$ meV, which is consistent with the calculation on the basis of Eqs. (3)-(5) giving the energy splitting of 11 meV (see supplementary materials).
We note that the lowest quasi-doublet should have a tiny gap due to the non-Kramers nature of a Tm$^{3+}$(4f$^{12}$) ion \cite{li2020partial,shen2019intertwined}, if one considers the effect of CEF potentials $\hat{O}^{\pm 3}_n$ ($n$ = 4, 6) in Eq. (1), which has remained unresolved in the present study.

Established the magnetic anisotropy originating from the CEF in $\textit{R}$AuGe, we focus on the magnetic transitions at low temperatures.
Figures 6(a)-(l) show the susceptibility at $\mu_0H = 0.1$ T as well as specific heat and thermal expansion at zero field near the transition temperatures.
Together with the structural parameters, we summarize the magnetic properties in Table~I.

Figures 6(a) and (b) compare the temperature dependences of physical properties of GdAuGe.
All the measurements were performed by using a unique single crystal.
There are anomalies in $M/H$ for both $H\parallel c$ and $H\perp c$ (Fig. 6(a)) indicating two successive magnetic transitions at $T_{\text{N}} =$ 16.9 K and $T_{\text{mag}}=$ 10.0 K.
In the specific heat measurement (Fig. 6(b)), on the other hand, we observed two anomalies at $T_{\text{N}} =$ 17.0 K and $T_{\text{HC}}=$ 14.8 K, reproducing the previous study \cite{gibson1996susceptibility}.
Note that the observed lattice-length change below $T_{\text{N}}$ can be mainly attributed to the exchange striction because the CEF anisotropy is absent \cite{doerr2005magnetostriction}.
No obvious anomaly at $T_{\text{mag}}$ or $T_{\text{HC}}$ potentially suggests that these transitions are related with spin-orientation change keeping $\bm{S}_i\cdot \bm{S}_j$ for local Gd spin moments ($\bm{S}_i$) unchanged.
Aside from the clear antiferromagnetic transition at 17 K, we observed secondary anomalies at different temperatures, $T_{\text{mag}}$ and $T_{\text{HC}}$, through different probes because each anomaly is too subtle to be seen through the other properties.
It is a possible future study to develop the magnetic phase diagram and to identify the evolution of the magnetic structures at low temperatures.

The magnetic phase transition of TbAuGe takes place at $T_{\text{N}} =$ 6 K as seen as a peak of $M/H$ for $H\perp c$ (Fig. 6(c)) and a tiny shoulder-type anomaly in $C_{\text{mag}}$ (Fig. 6(d)).
For $H\parallel c$, $M/H$ shows a kink at $T_{\text{N}}$ and is well suppressed in contrast to that for $H\perp c$ owing to the strong easy-plane type anisotropy.
Thermal expansion shows an anomaly at $T_{\text{N}}$.
We note that the lowest CEF level for  the non-Kramers Tb$^{3+}$ ion is a non-magnetic singlet isolated from the lowest excited state.
The energy scale of the separation is estimated to be 15 K on the basis of the simulation (see supplementary materials)).
Magnetic ordering in such a singlet-ground-state system is enabled by induced-moment instability upon the exchange interaction exceeding the CEF gap \cite{grover1965dynamical,goremychkin2008spin}.
This is consistent with the suppressed $C_{\text{mag}}$ peak at $T_{\text{N}}$ as observed in Pr-based compounds \cite{andres1972induced,kitazawa1998magnetic,anand2014investigations}. 
An additional anomaly associated with an incommensurate-commensurate transition is not identified down to $T=1.8$ K in our measurements.
A diffraction study is to be reported in future.

The antiferromagnetic transition in DyAuGe is clearer than that in TbAuGe as shown as a drop of $M/H$ at $T_{\text{N}}=$ 4 K (Fig. 6(e)).
For $H\perp c$, the $M$-$T$ curve shows a hysteresis at lower temperatures, suggesting that the in-plane magnetic field induces rearrangement of the magnetic domains breaking $C_6$ symmetry of the crystal.
This is consistent with a hysteresis also seen in the in-plane thermal expansion (Fig. 6(f)).
The $M/H$ shows a drop below $T_{\text{N}}$ for both $H\parallel c$ and $H\perp c$, suggesting that the magnetic moments can have both in- and out-of-plane components.
The temperature derivative $\text{d}M/\text{d}T$ for $H\perp c$ shows a peak clarifying $T_{\text{N}}$ (see the orange arrow in Fig. 6(e)).
The magnetic transition appears as a broad peak of specific heat (Fig. 6(f)) while the successive nature is not clearly resolved.

The magnetic susceptibility of HoAuGe is isotropic (with a subtle easy-plane nature) at high temperatures while an easy-axis nature is enhanced at low temperatures (Fig. 6(g)).
The antiferromagnetic transition can be seen as a weak drop of $M/H$ at $T_{\text{N1}}=5$ K and a tiny peak in $C_{\text{mag}}$ (black arrow in Fig. 6(h)).
The main peak in the specific heat is observed at $T_{\text{N2}}=4$ K, which is accompanied by a thermal-expansion jump and a kink in $M/H$ for $H\parallel c$.
The latter is more clearly seen in $\text{d}M/\text{d}T$ (orange arrow in Fig. 6(g)).
These anomalies are not discernible in $M/H$ for $H\perp c$, potentially indicating the predominant $c$-axis component of the spin moment in the magnetic ordered states.
The successive transitions support the previous neutron diffraction study, where a mixture of commensurate and incommensurate orders was identified below $T_{\text{N1}}=$ 6 K, which eventually developed to the pure commensurate order \cite{baran2000magnetic,baran2001neutron,gibson2001crystal}.
We do not observe multi-step anomalies at 5.6, 4, or 2 K in the specific heat measurement in contrast to a previous study \cite{gibson2001crystal} potentially because of the difference in sample quality or annealing condition.

ErAuGe shows an antiferromagnetic transition at $T_{\text{N}}= 3.2$ K.
$M$-$T$ curves (Fig. 6(i)) are hysteretic for both $H\parallel c$ and $H\perp c$ at low temperatures, suggesting a glassy nature of the magnetic order.
Large easy-axis type anisotropy indicates that the magnetic moments are aligned along the $c$ axis.
The transition can be clearly identified as a specific-heat peak and a thermal-expansion kink (Fig. 6(j)).
This is in contrast to the successive transitions at 3.7 K and 2.8 K associated with the incommensurate-commensurate transition observed in previous neutron diffraction \cite{baran2000magnetic,baran2001neutron}.

TmAuGe is the only ferromagnetic system among $\textit{R}$AuGe for $\textit{R}=$ Gd$-$Tm, as identified in Ref. [\onlinecite{kaczorowski2014magnetic}].
To see the divergence of magnetic susceptibility, we correct the demagnetization field for $H\parallel c$ (red curve in Fig. 6(k)).
Corrected magnetic susceptibility is obtained by $M/H_{\text{int}}$.
We fit the $M/H_{\text{int}}$ curve for $H\parallel c$ at low temperatures to the Curie-Weiss law ($\chi_{\parallel c}=\frac{C_{\text{low}}}{T-\mathit{\Theta}_{\parallel c\text{,low}}}$) as shown by the black curve in Fig. 6(k).
The Weiss temperature is $\mathit{\Theta} _{\parallel c\text{,low}}= 4.1$ K, consistent with $T_{\text{C}}$.
We obtain the effective moment $p_{\text{Ising}}=7.07 \mu_{\text{B}}$ by the formula for the Curie constant $C_{\text{low}}=N_{\text{A}}p_{\text{Ising}}^2/k_{\text{B}}$ in the case of the Ising-type magnetic moments.
The obtained $p_{\text{Ising}}$ agrees well with the expected value $p_{\text{Ising}}=g_JJ=7 \mu_{\text{B}}$ for the lowest quasi-doublet composed of $J_z=\pm 6$ well separated from the lowest excited state, which is consistent with the CEF analysis.

\begin{figure}[t]
	\includegraphics[width =  \columnwidth]{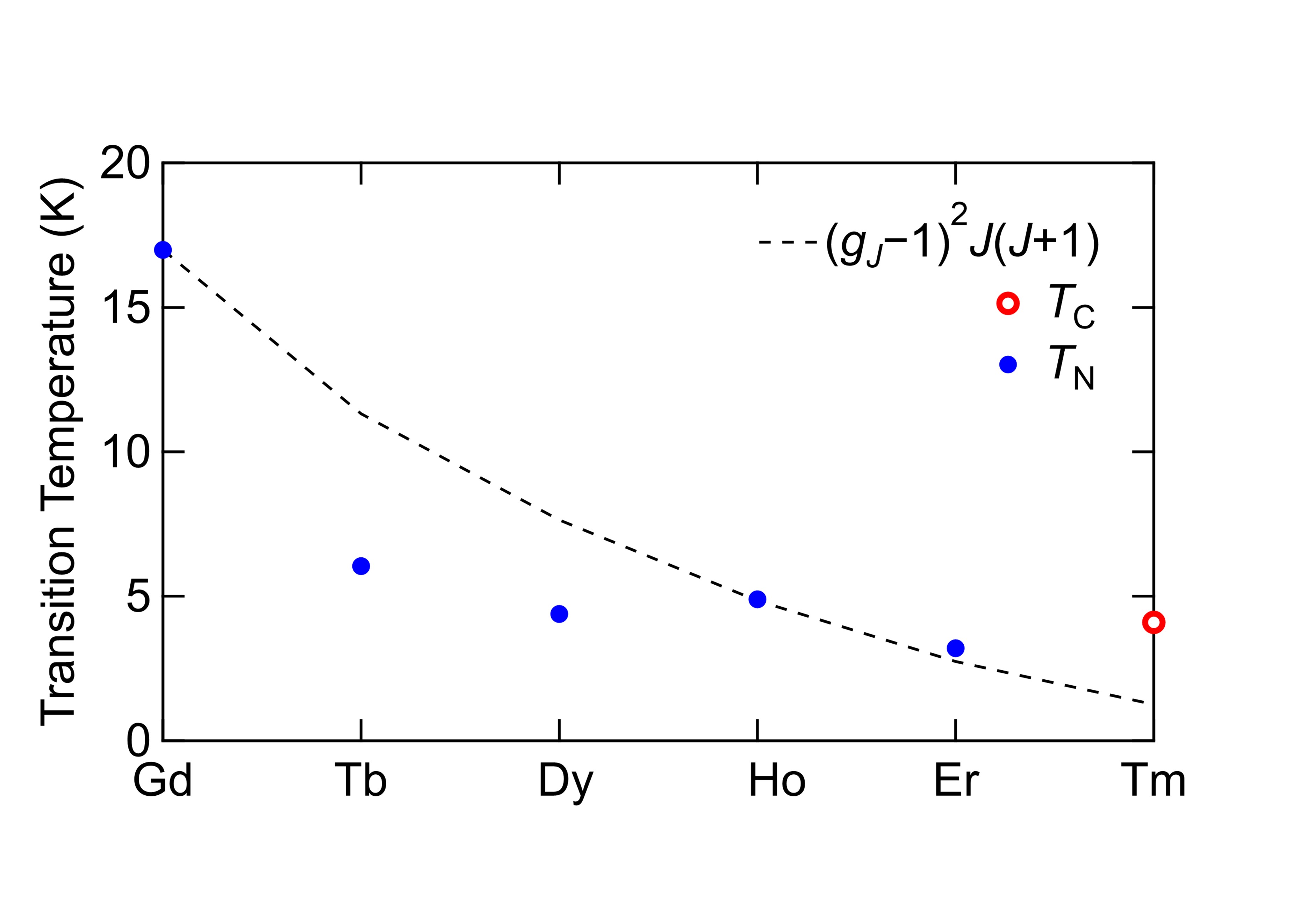}
	\caption{\label{fig7} $\textit{R}$ dependence of the transition temperatures.
	Blue (red) markers represent antiferromagnetic (ferromagnetic) transitions.
	Dashed line is the de Gennes factor $(g_{J}-1)^2J(J+1)$ normalized by the Gd value.
}
\end{figure}

Finally, we discuss the $\textit{R}$ dependence of magnetic transition temperature summarized in Fig. 7.
On the basis of the de Gennes scaling rule \cite{de1962indirect}, the transition temperatures among isostructural $\textit{R}$ compounds are supposed to be scaled with the de Gennes factor, $(g_J-1)^2J(J+1)$ (see the dashed curve in Fig. 7).
This argument is valid as long as the CEF effect and anisotropy of exchange interaction are negligible and Heisenberg-type RKKY interaction ($\mathcal{J}_{\text{ex}}(\bm{q}_{\text{mag}})$) stabilizes a magnetic order with the identical magnetic modulation vector $\bm{q}_{\text{mag}}$ among $\textit{R}$AuGe.
We observe the breakdown of the de Gennes scaling as it overestimates $T_{\text{N}}$ for Tb and Dy and underestimates $T_{\text{C}}$ for Tm.
Indeed, one of the above conditions is violated as the Tm compound is ferromagnetic ($\bm{q}_{\text{mag}}=(0, 0, 0)$) in contrast that the antiferromagnetic order with $\bm{q}_{\text{mag}}\sim (0.4-0.5, 0, 0)$ in the reciprocal lattice unit is favored for the other compounds \cite{gibson1998investigation,baran2000magnetic,baran2001neutron}.
We have to consider the modification of $\mathcal{J}_{\text{ex}}(\bm{q}_{\text{mag}})$, which is potentially sensitive to subtle details of the band structure in the present semimetallic materials.
We also note that the suppression of $T_{\text{N}}$ for $\textit{R}=$ Tb is reasonable because of the singlet ground state \cite{liu2003transition} while that of $\textit{R}=$ Dy is nontrivial.
Theories predict that the CEF effect is basically supposed to enhance the transition temperature in the case of the easy-axis type anisotropy ($B_{20}<0$, not valid for $\textit{R}=$ Dy) \cite{noakes1982effect,luong2001effect}, but the suppression of $T_{\text{N}}$ is possible when the easy-plane type anisotropy ($B_{20}>0$) surpasses the exchange interaction \cite{lines1975sensitivity}.
The observed thermal expansion ($\Delta L/L \sim 1.7 \times 10^{-4}$) in DyAuGe is the largest in the present $\textit{R}$AuGe family, which may reflect the magnetoelastic strain originating from the magnetic anisotropy rather than the exchange striction.
We might also have to consider the effect of magnetic frustration in DyAuGe to potentially promote the competition among noncollinear/noncoplanar magnetic structures suppressing $T_{\text{N}}$.

In summary, we succeeded in growing single crystals of $\textit{R}$AuGe for $\textit{R}=$ Y, Gd$-$Tm, and Lu and revealed the thermodynamic and magnetic properties.
Magnetic anisotropy is systematically determined by the CEF effect of $\textit{R}$ ions, and low-temperature transitions are dominated by the lowest CEF levels.
Continuous structural evolution of the polarity and its potential interplay with nontrivial magnetism would be viewed as a promising platform for studying metallic multiferroics \cite{hickox2022polar,bhowal2022polar}. 
Findings in the present study provide basic understanding of single crystalline properties of a family of polar semimetal $\textit{R}$AuGe.
Transport properties intertwinned with frustrated magnetism is a desired future study, which will be reported elsewhere.

\begin{acknowledgements}
T.K. was financially supported by Ministry of Education Culture Sports Science and Technology (MEXT) Leading Initiative for Excellent Young Researchers (JPMXS0320200135), Japan Society for the Promotion of Science (JSPS) KAKENHI Grant-in-Aid for Young Scientists B (No. 21K13874).
M.G. was supported by the JSPS KAKENHI Grant-in-Aid for Scientific Research (No. 20J10988).
S.K. was supported by JSPS KAKENHI Grant-in-Aid for Early-Career Scientists (No. 22K14010).
This work was partly supported by JSPS KAKENHI Grant-in-Aid for Scientific Research on Innovative Areas Quantum Liquid Crystals (No. JP19H05826 and No. 19H01835).
The neutron experiment at the Materials and Life Science Experimental Facility of the J-PARC was performed under a user program (Proposal No. 2022A0045).
This work was partly performed using the facilities of the Materials Design and Characterization Laboratory in the Institute for Solid State Physics, the University of Tokyo.
S.K. thanks K. Adachi and D. Hashizume for the support of the in-house XRD experiments.
\end{acknowledgements}
\bibliography{reference}

%merlin.mbs apsrev4-1.bst 2010-07-25 4.21a (PWD, AO, DPC) hacked
%Control: key (0)
%Control: author (0) dotless jnrlst
%Control: editor formatted (1) identically to author
%Control: production of article title (0) allowed
%Control: page (1) range
%Control: year (0) verbatim
%Control: production of eprint (0) enabled
\begin{thebibliography}{63}%
\makeatletter
\providecommand \@ifxundefined [1]{%
 \@ifx{#1\undefined}
}%
\providecommand \@ifnum [1]{%
 \ifnum #1\expandafter \@firstoftwo
 \else \expandafter \@secondoftwo
 \fi
}%
\providecommand \@ifx [1]{%
 \ifx #1\expandafter \@firstoftwo
 \else \expandafter \@secondoftwo
 \fi
}%
\providecommand \natexlab [1]{#1}%
\providecommand \enquote  [1]{``#1''}%
\providecommand \bibnamefont  [1]{#1}%
\providecommand \bibfnamefont [1]{#1}%
\providecommand \citenamefont [1]{#1}%
\providecommand \href@noop [0]{\@secondoftwo}%
\providecommand \href [0]{\begingroup \@sanitize@url \@href}%
\providecommand \@href[1]{\@@startlink{#1}\@@href}%
\providecommand \@@href[1]{\endgroup#1\@@endlink}%
\providecommand \@sanitize@url [0]{\catcode `\\12\catcode `\$12\catcode
  `\&12\catcode `\#12\catcode `\^12\catcode `\_12\catcode `\%12\relax}%
\providecommand \@@startlink[1]{}%
\providecommand \@@endlink[0]{}%
\providecommand \url  [0]{\begingroup\@sanitize@url \@url }%
\providecommand \@url [1]{\endgroup\@href {#1}{\urlprefix }}%
\providecommand \urlprefix  [0]{URL }%
\providecommand \Eprint [0]{\href }%
\providecommand \doibase [0]{http://dx.doi.org/}%
\providecommand \selectlanguage [0]{\@gobble}%
\providecommand \bibinfo  [0]{\@secondoftwo}%
\providecommand \bibfield  [0]{\@secondoftwo}%
\providecommand \translation [1]{[#1]}%
\providecommand \BibitemOpen [0]{}%
\providecommand \bibitemStop [0]{}%
\providecommand \bibitemNoStop [0]{.\EOS\space}%
\providecommand \EOS [0]{\spacefactor3000\relax}%
\providecommand \BibitemShut  [1]{\csname bibitem#1\endcsname}%
\let\auto@bib@innerbib\@empty
%</preamble>
\bibitem [{\citenamefont {Shi}\ \emph {et~al.}(2013)\citenamefont {Shi},
  \citenamefont {Guo}, \citenamefont {Wang}, \citenamefont {Princep},
  \citenamefont {Khalyavin}, \citenamefont {Manuel}, \citenamefont {Michiue},
  \citenamefont {Sato}, \citenamefont {Tsuda}, \citenamefont {Yu} \emph
  {et~al.}}]{shi2013ferroelectric}%
  \BibitemOpen
  \bibfield  {author} {\bibinfo {author} {\bibfnamefont {Y.}~\bibnamefont
  {Shi}}, \bibinfo {author} {\bibfnamefont {Y.}~\bibnamefont {Guo}}, \bibinfo
  {author} {\bibfnamefont {X.}~\bibnamefont {Wang}}, \bibinfo {author}
  {\bibfnamefont {A.~J.}\ \bibnamefont {Princep}}, \bibinfo {author}
  {\bibfnamefont {D.}~\bibnamefont {Khalyavin}}, \bibinfo {author}
  {\bibfnamefont {P.}~\bibnamefont {Manuel}}, \bibinfo {author} {\bibfnamefont
  {Y.}~\bibnamefont {Michiue}}, \bibinfo {author} {\bibfnamefont
  {A.}~\bibnamefont {Sato}}, \bibinfo {author} {\bibfnamefont {K.}~\bibnamefont
  {Tsuda}}, \bibinfo {author} {\bibfnamefont {S.}~\bibnamefont {Yu}},  \emph
  {et~al.},\ }\bibfield  {title} {\enquote {\bibinfo {title} {A
  ferroelectric-like structural transition in a metal},}\ }\href@noop {}
  {\bibfield  {journal} {\bibinfo  {journal} {Nat. Mater.}\ }\textbf {\bibinfo
  {volume} {12}},\ \bibinfo {pages} {1024--1027} (\bibinfo {year}
  {2013})}\BibitemShut {NoStop}%
\bibitem [{\citenamefont {Benedek}\ and\ \citenamefont
  {Birol}(2016)}]{benedek2016ferroelectric}%
  \BibitemOpen
  \bibfield  {author} {\bibinfo {author} {\bibfnamefont {N.~A.}\ \bibnamefont
  {Benedek}}\ and\ \bibinfo {author} {\bibfnamefont {T.}~\bibnamefont
  {Birol}},\ }\bibfield  {title} {\enquote {\bibinfo {title}
  {‘{F}erroelectric’ metals reexamined: fundamental mechanisms and design
  considerations for new materials},}\ }\href@noop {} {\bibfield  {journal}
  {\bibinfo  {journal} {J. Mater. Chem. C}\ }\textbf {\bibinfo {volume} {4}},\
  \bibinfo {pages} {4000} (\bibinfo {year} {2016})}\BibitemShut {NoStop}%
\bibitem [{\citenamefont {Garrity}\ \emph {et~al.}(2014)\citenamefont
  {Garrity}, \citenamefont {Rabe},\ and\ \citenamefont
  {Vanderbilt}}]{garrity2014hyperferroelectrics}%
  \BibitemOpen
  \bibfield  {author} {\bibinfo {author} {\bibfnamefont {K.~F.}\ \bibnamefont
  {Garrity}}, \bibinfo {author} {\bibfnamefont {K.~M.}\ \bibnamefont {Rabe}}, \
  and\ \bibinfo {author} {\bibfnamefont {D.}~\bibnamefont {Vanderbilt}},\
  }\bibfield  {title} {\enquote {\bibinfo {title} {Hyperferroelectrics: proper
  ferroelectrics with persistent polarization},}\ }\href@noop {} {\bibfield
  {journal} {\bibinfo  {journal} {Phys. Rev. Lett.}\ }\textbf {\bibinfo
  {volume} {112}},\ \bibinfo {pages} {127601} (\bibinfo {year}
  {2014})}\BibitemShut {NoStop}%
\bibitem [{\citenamefont {Zhou}\ and\ \citenamefont
  {Ariando}(2020)}]{zhou2020review}%
  \BibitemOpen
  \bibfield  {author} {\bibinfo {author} {\bibfnamefont {W.~X.}\ \bibnamefont
  {Zhou}}\ and\ \bibinfo {author} {\bibfnamefont {A.}~\bibnamefont {Ariando}},\
  }\bibfield  {title} {\enquote {\bibinfo {title} {Review on
  ferroelectric/polar metals},}\ }\href@noop {} {\bibfield  {journal} {\bibinfo
   {journal} {Jpn. J. Appl. Phys.}\ }\textbf {\bibinfo {volume} {59}},\
  \bibinfo {pages} {SI0802} (\bibinfo {year} {2020})}\BibitemShut {NoStop}%
\bibitem [{\citenamefont {Narayan}(2015)}]{narayan2015class}%
  \BibitemOpen
  \bibfield  {author} {\bibinfo {author} {\bibfnamefont {A.}~\bibnamefont
  {Narayan}},\ }\bibfield  {title} {\enquote {\bibinfo {title} {Class of
  {R}ashba ferroelectrics in hexagonal semiconductors},}\ }\href@noop {}
  {\bibfield  {journal} {\bibinfo  {journal} {Phys. Rev. B}\ }\textbf {\bibinfo
  {volume} {92}},\ \bibinfo {pages} {220101} (\bibinfo {year}
  {2015})}\BibitemShut {NoStop}%
\bibitem [{\citenamefont {Fei}\ \emph {et~al.}(2018)\citenamefont {Fei},
  \citenamefont {Zhao}, \citenamefont {Palomaki}, \citenamefont {Sun},
  \citenamefont {Miller}, \citenamefont {Zhao}, \citenamefont {Yan},
  \citenamefont {Xu},\ and\ \citenamefont {Cobden}}]{fei2018ferroelectric}%
  \BibitemOpen
  \bibfield  {author} {\bibinfo {author} {\bibfnamefont {Z.}~\bibnamefont
  {Fei}}, \bibinfo {author} {\bibfnamefont {W.}~\bibnamefont {Zhao}}, \bibinfo
  {author} {\bibfnamefont {T.~A.}\ \bibnamefont {Palomaki}}, \bibinfo {author}
  {\bibfnamefont {B.}~\bibnamefont {Sun}}, \bibinfo {author} {\bibfnamefont
  {M.~K.}\ \bibnamefont {Miller}}, \bibinfo {author} {\bibfnamefont
  {Z.}~\bibnamefont {Zhao}}, \bibinfo {author} {\bibfnamefont {J.}~\bibnamefont
  {Yan}}, \bibinfo {author} {\bibfnamefont {X.}~\bibnamefont {Xu}}, \ and\
  \bibinfo {author} {\bibfnamefont {D.~H.}\ \bibnamefont {Cobden}},\ }\bibfield
   {title} {\enquote {\bibinfo {title} {Ferroelectric switching of a
  two-dimensional metal},}\ }\href@noop {} {\bibfield  {journal} {\bibinfo
  {journal} {Nature}\ }\textbf {\bibinfo {volume} {560}},\ \bibinfo {pages}
  {336--339} (\bibinfo {year} {2018})}\BibitemShut {NoStop}%
\bibitem [{\citenamefont {Cao}\ \emph {et~al.}(2017)\citenamefont {Cao},
  \citenamefont {Tang}, \citenamefont {Xu}, \citenamefont {Wu}, \citenamefont
  {Gu},\ and\ \citenamefont {Duan}}]{cao2017dirac}%
  \BibitemOpen
  \bibfield  {author} {\bibinfo {author} {\bibfnamefont {W.}~\bibnamefont
  {Cao}}, \bibinfo {author} {\bibfnamefont {P.}~\bibnamefont {Tang}}, \bibinfo
  {author} {\bibfnamefont {Y.}~\bibnamefont {Xu}}, \bibinfo {author}
  {\bibfnamefont {J.}~\bibnamefont {Wu}}, \bibinfo {author} {\bibfnamefont
  {B.-L.}\ \bibnamefont {Gu}}, \ and\ \bibinfo {author} {\bibfnamefont
  {W.}~\bibnamefont {Duan}},\ }\bibfield  {title} {\enquote {\bibinfo {title}
  {Dirac semimetal phase in hexagonal \ce{LiZnBi}},}\ }\href@noop {} {\bibfield
   {journal} {\bibinfo  {journal} {Phys. Rev. B}\ }\textbf {\bibinfo {volume}
  {96}},\ \bibinfo {pages} {115203} (\bibinfo {year} {2017})}\BibitemShut
  {NoStop}%
\bibitem [{\citenamefont {Gao}\ \emph {et~al.}(2018)\citenamefont {Gao},
  \citenamefont {Kim}, \citenamefont {Venderbos}, \citenamefont {Kane},
  \citenamefont {Mele}, \citenamefont {Rappe},\ and\ \citenamefont
  {Ren}}]{gao2018dirac}%
  \BibitemOpen
  \bibfield  {author} {\bibinfo {author} {\bibfnamefont {H.}~\bibnamefont
  {Gao}}, \bibinfo {author} {\bibfnamefont {Y.}~\bibnamefont {Kim}}, \bibinfo
  {author} {\bibfnamefont {J.~W.~F.}\ \bibnamefont {Venderbos}}, \bibinfo
  {author} {\bibfnamefont {C.~L.}\ \bibnamefont {Kane}}, \bibinfo {author}
  {\bibfnamefont {E.~J.}\ \bibnamefont {Mele}}, \bibinfo {author}
  {\bibfnamefont {A.~M}\ \bibnamefont {Rappe}}, \ and\ \bibinfo {author}
  {\bibfnamefont {W.}~\bibnamefont {Ren}},\ }\bibfield  {title} {\enquote
  {\bibinfo {title} {Dirac-{W}eyl semimetal: Coexistence of {D}irac and {W}eyl
  fermions in polar hexagonal $\textit{A B C}$ crystals},}\ }\href@noop {}
  {\bibfield  {journal} {\bibinfo  {journal} {Phys. Rev. Lett.}\ }\textbf
  {\bibinfo {volume} {121}},\ \bibinfo {pages} {106404} (\bibinfo {year}
  {2018})}\BibitemShut {NoStop}%
\bibitem [{\citenamefont {Gao}\ \emph {et~al.}(2021)\citenamefont {Gao},
  \citenamefont {Strockoz}, \citenamefont {Frakulla}, \citenamefont
  {Venderbos},\ and\ \citenamefont {Weng}}]{gao2021noncentrosymmetric}%
  \BibitemOpen
  \bibfield  {author} {\bibinfo {author} {\bibfnamefont {H.}~\bibnamefont
  {Gao}}, \bibinfo {author} {\bibfnamefont {J.}~\bibnamefont {Strockoz}},
  \bibinfo {author} {\bibfnamefont {M.}~\bibnamefont {Frakulla}}, \bibinfo
  {author} {\bibfnamefont {J.~W.~F.}\ \bibnamefont {Venderbos}}, \ and\
  \bibinfo {author} {\bibfnamefont {H.}~\bibnamefont {Weng}},\ }\bibfield
  {title} {\enquote {\bibinfo {title} {Noncentrosymmetric topological {D}irac
  semimetals in three dimensions},}\ }\href@noop {} {\bibfield  {journal}
  {\bibinfo  {journal} {Phys. Rev. B}\ }\textbf {\bibinfo {volume} {103}},\
  \bibinfo {pages} {205151} (\bibinfo {year} {2021})}\BibitemShut {NoStop}%
\bibitem [{\citenamefont {Du}\ \emph {et~al.}(2015)\citenamefont {Du},
  \citenamefont {Wan}, \citenamefont {Wang}, \citenamefont {Sheng},
  \citenamefont {Duan},\ and\ \citenamefont {Wan}}]{du2015dirac}%
  \BibitemOpen
  \bibfield  {author} {\bibinfo {author} {\bibfnamefont {Y.}~\bibnamefont
  {Du}}, \bibinfo {author} {\bibfnamefont {B.}~\bibnamefont {Wan}}, \bibinfo
  {author} {\bibfnamefont {D.}~\bibnamefont {Wang}}, \bibinfo {author}
  {\bibfnamefont {L.}~\bibnamefont {Sheng}}, \bibinfo {author} {\bibfnamefont
  {C.-G.}\ \bibnamefont {Duan}}, \ and\ \bibinfo {author} {\bibfnamefont
  {X.}~\bibnamefont {Wan}},\ }\bibfield  {title} {\enquote {\bibinfo {title}
  {Dirac and {w}eyl semimetal in $\textit{XY}${B}i ($\textit{X}=$ {B}a, {E}u;
  $\textit{Y}=$ {C}u, {A}g and {A}u)},}\ }\href@noop {} {\bibfield  {journal}
  {\bibinfo  {journal} {Sci. Rep.}\ }\textbf {\bibinfo {volume} {5}},\ \bibinfo
  {pages} {1--8} (\bibinfo {year} {2015})}\BibitemShut {NoStop}%
\bibitem [{\citenamefont {Mondal}\ \emph {et~al.}(2019)\citenamefont {Mondal},
  \citenamefont {Barman}, \citenamefont {Alam},\ and\ \citenamefont
  {Pathak}}]{mondal2019broken}%
  \BibitemOpen
  \bibfield  {author} {\bibinfo {author} {\bibfnamefont {C.}~\bibnamefont
  {Mondal}}, \bibinfo {author} {\bibfnamefont {C.~K.}\ \bibnamefont {Barman}},
  \bibinfo {author} {\bibfnamefont {A.}~\bibnamefont {Alam}}, \ and\ \bibinfo
  {author} {\bibfnamefont {B.}~\bibnamefont {Pathak}},\ }\bibfield  {title}
  {\enquote {\bibinfo {title} {Broken symmetry driven phase transitions from a
  topological semimetal to a gapped topological phase in \ce{SrAgAs}},}\
  }\href@noop {} {\bibfield  {journal} {\bibinfo  {journal} {Phys. Rev. B}\
  }\textbf {\bibinfo {volume} {99}},\ \bibinfo {pages} {205112} (\bibinfo
  {year} {2019})}\BibitemShut {NoStop}%
\bibitem [{\citenamefont {Gaudet}\ \emph {et~al.}(2021)\citenamefont {Gaudet},
  \citenamefont {Yang}, \citenamefont {Baidya}, \citenamefont {Lu},
  \citenamefont {Xu}, \citenamefont {Zhao}, \citenamefont {Rodriguez-Rivera},
  \citenamefont {Hoffmann}, \citenamefont {Graf}, \citenamefont {Torchinsky}
  \emph {et~al.}}]{gaudet2021weyl}%
  \BibitemOpen
  \bibfield  {author} {\bibinfo {author} {\bibfnamefont {J.}~\bibnamefont
  {Gaudet}}, \bibinfo {author} {\bibfnamefont {H.-Y.}\ \bibnamefont {Yang}},
  \bibinfo {author} {\bibfnamefont {S.}~\bibnamefont {Baidya}}, \bibinfo
  {author} {\bibfnamefont {B.}~\bibnamefont {Lu}}, \bibinfo {author}
  {\bibfnamefont {G.}~\bibnamefont {Xu}}, \bibinfo {author} {\bibfnamefont
  {Y.}~\bibnamefont {Zhao}}, \bibinfo {author} {\bibfnamefont {J.~A.}\
  \bibnamefont {Rodriguez-Rivera}}, \bibinfo {author} {\bibfnamefont {C.~M.}\
  \bibnamefont {Hoffmann}}, \bibinfo {author} {\bibfnamefont {D.~E.}\
  \bibnamefont {Graf}}, \bibinfo {author} {\bibfnamefont {D.~H.}\ \bibnamefont
  {Torchinsky}},  \emph {et~al.},\ }\bibfield  {title} {\enquote {\bibinfo
  {title} {Weyl-mediated helical magnetism in \ce{NdAlSi}},}\ }\href@noop {}
  {\bibfield  {journal} {\bibinfo  {journal} {Nat. Mater.}\ }\textbf {\bibinfo
  {volume} {20}},\ \bibinfo {pages} {1650--1656} (\bibinfo {year}
  {2021})}\BibitemShut {NoStop}%
\bibitem [{\citenamefont {Srivastava}\ \emph {et~al.}(2020)\citenamefont
  {Srivastava}, \citenamefont {Devi}, \citenamefont {Sharma}, \citenamefont
  {Ma}, \citenamefont {Deniz}, \citenamefont {Meyerheim}, \citenamefont
  {Felser},\ and\ \citenamefont {Parkin}}]{srivastava2020observation}%
  \BibitemOpen
  \bibfield  {author} {\bibinfo {author} {\bibfnamefont {A.~K.}\ \bibnamefont
  {Srivastava}}, \bibinfo {author} {\bibfnamefont {P.}~\bibnamefont {Devi}},
  \bibinfo {author} {\bibfnamefont {A.~K.}\ \bibnamefont {Sharma}}, \bibinfo
  {author} {\bibfnamefont {T.}~\bibnamefont {Ma}}, \bibinfo {author}
  {\bibfnamefont {H.}~\bibnamefont {Deniz}}, \bibinfo {author} {\bibfnamefont
  {H.~L.}\ \bibnamefont {Meyerheim}}, \bibinfo {author} {\bibfnamefont
  {C.}~\bibnamefont {Felser}}, \ and\ \bibinfo {author} {\bibfnamefont
  {S.~S.~P.}\ \bibnamefont {Parkin}},\ }\bibfield  {title} {\enquote {\bibinfo
  {title} {Observation of robust {N}\'{e}el skyrmions in metallic
  \ce{PtMnGa}},}\ }\href@noop {} {\bibfield  {journal} {\bibinfo  {journal}
  {Adv. Mater.}\ }\textbf {\bibinfo {volume} {32}},\ \bibinfo {pages} {1904327}
  (\bibinfo {year} {2020})}\BibitemShut {NoStop}%
\bibitem [{\citenamefont {Zhang}\ \emph
  {et~al.}(2022{\natexlab{a}})\citenamefont {Zhang}, \citenamefont {Raftrey},
  \citenamefont {Chan}, \citenamefont {Shao}, \citenamefont {Chen},
  \citenamefont {Chen}, \citenamefont {Huang}, \citenamefont {Reichanadter},
  \citenamefont {Dong}, \citenamefont {Susarla} \emph
  {et~al.}}]{zhang2022room}%
  \BibitemOpen
  \bibfield  {author} {\bibinfo {author} {\bibfnamefont {H.}~\bibnamefont
  {Zhang}}, \bibinfo {author} {\bibfnamefont {D.}~\bibnamefont {Raftrey}},
  \bibinfo {author} {\bibfnamefont {Y.-T.}\ \bibnamefont {Chan}}, \bibinfo
  {author} {\bibfnamefont {Y.-T.}\ \bibnamefont {Shao}}, \bibinfo {author}
  {\bibfnamefont {R.}~\bibnamefont {Chen}}, \bibinfo {author} {\bibfnamefont
  {X.}~\bibnamefont {Chen}}, \bibinfo {author} {\bibfnamefont {X.}~\bibnamefont
  {Huang}}, \bibinfo {author} {\bibfnamefont {J.~T.}\ \bibnamefont
  {Reichanadter}}, \bibinfo {author} {\bibfnamefont {K.}~\bibnamefont {Dong}},
  \bibinfo {author} {\bibfnamefont {S.}~\bibnamefont {Susarla}},  \emph
  {et~al.},\ }\bibfield  {title} {\enquote {\bibinfo {title} {Room-temperature
  skyrmion lattice in a layered magnet
  ({F}e$_{0.5}${C}o$_{0.5}$)$_{5}${G}e{T}e$_{2}$},}\ }\href@noop {} {\bibfield
  {journal} {\bibinfo  {journal} {Sci. Adv.}\ }\textbf {\bibinfo {volume}
  {8}},\ \bibinfo {pages} {eabm7103} (\bibinfo {year}
  {2022}{\natexlab{a}})}\BibitemShut {NoStop}%
\bibitem [{\citenamefont {Zhang}\ \emph
  {et~al.}(2022{\natexlab{b}})\citenamefont {Zhang}, \citenamefont {Shao},
  \citenamefont {Chen}, \citenamefont {Chen}, \citenamefont {Susarla},
  \citenamefont {Raftrey}, \citenamefont {Reichanadter}, \citenamefont
  {Caretta}, \citenamefont {Huang}, \citenamefont {Settineri} \emph
  {et~al.}}]{zhang2022room2}%
  \BibitemOpen
  \bibfield  {author} {\bibinfo {author} {\bibfnamefont {H.}~\bibnamefont
  {Zhang}}, \bibinfo {author} {\bibfnamefont {Y.-T.}\ \bibnamefont {Shao}},
  \bibinfo {author} {\bibfnamefont {R.}~\bibnamefont {Chen}}, \bibinfo {author}
  {\bibfnamefont {X.}~\bibnamefont {Chen}}, \bibinfo {author} {\bibfnamefont
  {S.}~\bibnamefont {Susarla}}, \bibinfo {author} {\bibfnamefont
  {D.}~\bibnamefont {Raftrey}}, \bibinfo {author} {\bibfnamefont {J.~T.}\
  \bibnamefont {Reichanadter}}, \bibinfo {author} {\bibfnamefont
  {L.}~\bibnamefont {Caretta}}, \bibinfo {author} {\bibfnamefont
  {X.}~\bibnamefont {Huang}}, \bibinfo {author} {\bibfnamefont {N.~S.}\
  \bibnamefont {Settineri}},  \emph {et~al.},\ }\bibfield  {title} {\enquote
  {\bibinfo {title} {A room temperature polar magnetic metal},}\ }\href@noop {}
  {\bibfield  {journal} {\bibinfo  {journal} {Phys. Rev. Mater.}\ }\textbf
  {\bibinfo {volume} {6}},\ \bibinfo {pages} {044403} (\bibinfo {year}
  {2022}{\natexlab{b}})}\BibitemShut {NoStop}%
\bibitem [{\citenamefont {Gupta}\ and\ \citenamefont
  {Suresh}(2015)}]{gupta2015review}%
  \BibitemOpen
  \bibfield  {author} {\bibinfo {author} {\bibfnamefont {S.}~\bibnamefont
  {Gupta}}\ and\ \bibinfo {author} {\bibfnamefont {K.~G.}\ \bibnamefont
  {Suresh}},\ }\bibfield  {title} {\enquote {\bibinfo {title} {Review on
  magnetic and related properties of $\textit{RTX}$ compounds},}\ }\href@noop
  {} {\bibfield  {journal} {\bibinfo  {journal} {J. Alloys Compd.}\ }\textbf
  {\bibinfo {volume} {618}},\ \bibinfo {pages} {562} (\bibinfo {year}
  {2015})}\BibitemShut {NoStop}%
\bibitem [{\citenamefont {Rossi}\ \emph {et~al.}(1992)\citenamefont {Rossi},
  \citenamefont {Marazza},\ and\ \citenamefont {Ferro}}]{rossi1992ternary}%
  \BibitemOpen
  \bibfield  {author} {\bibinfo {author} {\bibfnamefont {D.}~\bibnamefont
  {Rossi}}, \bibinfo {author} {\bibfnamefont {R.}~\bibnamefont {Marazza}}, \
  and\ \bibinfo {author} {\bibfnamefont {R.}~\bibnamefont {Ferro}},\ }\bibfield
   {title} {\enquote {\bibinfo {title} {Ternary rare earth alloys:
  $\textit{R}${A}u{G}e compounds},}\ }\href@noop {} {\bibfield  {journal}
  {\bibinfo  {journal} {J. Alloys Compd.}\ }\textbf {\bibinfo {volume} {187}},\
  \bibinfo {pages} {267} (\bibinfo {year} {1992})}\BibitemShut {NoStop}%
\bibitem [{\citenamefont {P\"{o}ttgen}\ \emph
  {et~al.}(1996{\natexlab{a}})\citenamefont {P\"{o}ttgen}, \citenamefont
  {Borrmann}, \citenamefont {Felser}, \citenamefont {Jepsen}, \citenamefont
  {Henn}, \citenamefont {Kremer},\ and\ \citenamefont
  {Simon}}]{pottgen1996crystal}%
  \BibitemOpen
  \bibfield  {author} {\bibinfo {author} {\bibfnamefont {R.}~\bibnamefont
  {P\"{o}ttgen}}, \bibinfo {author} {\bibfnamefont {H.}~\bibnamefont
  {Borrmann}}, \bibinfo {author} {\bibfnamefont {C.}~\bibnamefont {Felser}},
  \bibinfo {author} {\bibfnamefont {O.}~\bibnamefont {Jepsen}}, \bibinfo
  {author} {\bibfnamefont {R.}~\bibnamefont {Henn}}, \bibinfo {author}
  {\bibfnamefont {R.~K.}\ \bibnamefont {Kremer}}, \ and\ \bibinfo {author}
  {\bibfnamefont {A.}~\bibnamefont {Simon}},\ }\bibfield  {title} {\enquote
  {\bibinfo {title} {Crystal and electronic structures of {S}c{A}u{G}e,
  {C}e{A}u{G}e, and {L}u{A}u{G}e: a transition from two- to three-dimensional
  [{A}u{G}e] polyanions},}\ }\href@noop {} {\bibfield  {journal} {\bibinfo
  {journal} {J. Alloys Compd.}\ }\textbf {\bibinfo {volume} {235}},\ \bibinfo
  {pages} {170} (\bibinfo {year} {1996}{\natexlab{a}})}\BibitemShut {NoStop}%
\bibitem [{\citenamefont {Bennett}\ \emph {et~al.}(2012)\citenamefont
  {Bennett}, \citenamefont {Garrity}, \citenamefont {Rabe},\ and\ \citenamefont
  {Vanderbilt}}]{bennett2012hexagonal}%
  \BibitemOpen
  \bibfield  {author} {\bibinfo {author} {\bibfnamefont {J.~W.}\ \bibnamefont
  {Bennett}}, \bibinfo {author} {\bibfnamefont {K.~F.}\ \bibnamefont
  {Garrity}}, \bibinfo {author} {\bibfnamefont {K.~M.}\ \bibnamefont {Rabe}}, \
  and\ \bibinfo {author} {\bibfnamefont {D.}~\bibnamefont {Vanderbilt}},\
  }\bibfield  {title} {\enquote {\bibinfo {title} {Hexagonal $\textit{ABC}$
  semiconductors as ferroelectrics},}\ }\href@noop {} {\bibfield  {journal}
  {\bibinfo  {journal} {Phys. Rev. Lett.}\ }\textbf {\bibinfo {volume} {109}},\
  \bibinfo {pages} {167602} (\bibinfo {year} {2012})}\BibitemShut {NoStop}%
\bibitem [{\citenamefont {Gibson}\ \emph {et~al.}(2015)\citenamefont {Gibson},
  \citenamefont {Schoop}, \citenamefont {Muechler}, \citenamefont {Xie},
  \citenamefont {Hirschberger}, \citenamefont {Ong}, \citenamefont {Car},\ and\
  \citenamefont {Cava}}]{gibson2015three}%
  \BibitemOpen
  \bibfield  {author} {\bibinfo {author} {\bibfnamefont {Q.~D.}\ \bibnamefont
  {Gibson}}, \bibinfo {author} {\bibfnamefont {L.~M.}\ \bibnamefont {Schoop}},
  \bibinfo {author} {\bibfnamefont {L.}~\bibnamefont {Muechler}}, \bibinfo
  {author} {\bibfnamefont {L.~S.}\ \bibnamefont {Xie}}, \bibinfo {author}
  {\bibfnamefont {M.}~\bibnamefont {Hirschberger}}, \bibinfo {author}
  {\bibfnamefont {N.~P.}\ \bibnamefont {Ong}}, \bibinfo {author} {\bibfnamefont
  {R.}~\bibnamefont {Car}}, \ and\ \bibinfo {author} {\bibfnamefont {R.~J.}\
  \bibnamefont {Cava}},\ }\bibfield  {title} {\enquote {\bibinfo {title}
  {Three-dimensional {D}irac semimetals: design principles and predictions of
  new materials},}\ }\href@noop {} {\bibfield  {journal} {\bibinfo  {journal}
  {Phys. Rev. B}\ }\textbf {\bibinfo {volume} {91}},\ \bibinfo {pages} {205128}
  (\bibinfo {year} {2015})}\BibitemShut {NoStop}%
\bibitem [{\citenamefont {Di~Sante}\ \emph {et~al.}(2016)\citenamefont
  {Di~Sante}, \citenamefont {Barone}, \citenamefont {Stroppa}, \citenamefont
  {Garrity}, \citenamefont {Vanderbilt},\ and\ \citenamefont
  {Picozzi}}]{di2016intertwined}%
  \BibitemOpen
  \bibfield  {author} {\bibinfo {author} {\bibfnamefont {D.}~\bibnamefont
  {Di~Sante}}, \bibinfo {author} {\bibfnamefont {P.}~\bibnamefont {Barone}},
  \bibinfo {author} {\bibfnamefont {A.}~\bibnamefont {Stroppa}}, \bibinfo
  {author} {\bibfnamefont {K.~F.}\ \bibnamefont {Garrity}}, \bibinfo {author}
  {\bibfnamefont {D.}~\bibnamefont {Vanderbilt}}, \ and\ \bibinfo {author}
  {\bibfnamefont {S.}~\bibnamefont {Picozzi}},\ }\bibfield  {title} {\enquote
  {\bibinfo {title} {Intertwined {R}ashba, {D}irac, and {W}eyl fermions in
  hexagonal hyperferroelectrics},}\ }\href@noop {} {\bibfield  {journal}
  {\bibinfo  {journal} {Phys. Rev. Lett.}\ }\textbf {\bibinfo {volume} {117}},\
  \bibinfo {pages} {076401} (\bibinfo {year} {2016})}\BibitemShut {NoStop}%
\bibitem [{\citenamefont {Zhang}\ \emph {et~al.}(2019)\citenamefont {Zhang},
  \citenamefont {Huang}, \citenamefont {Mei},\ and\ \citenamefont
  {Shi}}]{zhang2019influences}%
  \BibitemOpen
  \bibfield  {author} {\bibinfo {author} {\bibfnamefont {H.}~\bibnamefont
  {Zhang}}, \bibinfo {author} {\bibfnamefont {W.}~\bibnamefont {Huang}},
  \bibinfo {author} {\bibfnamefont {J.-W.}\ \bibnamefont {Mei}}, \ and\
  \bibinfo {author} {\bibfnamefont {X.-Q.}\ \bibnamefont {Shi}},\ }\bibfield
  {title} {\enquote {\bibinfo {title} {Influences of spin-orbit coupling on
  {F}ermi surfaces and {D}irac cones in ferroelectriclike polar metals},}\
  }\href@noop {} {\bibfield  {journal} {\bibinfo  {journal} {Phys. Rev. B}\
  }\textbf {\bibinfo {volume} {99}},\ \bibinfo {pages} {195154} (\bibinfo
  {year} {2019})}\BibitemShut {NoStop}%
\bibitem [{\citenamefont {P\"{o}ttgen}\ \emph
  {et~al.}(1996{\natexlab{b}})\citenamefont {P\"{o}ttgen}, \citenamefont
  {Borrmann},\ and\ \citenamefont {Kremer}}]{pottgen1996ferromagnetic}%
  \BibitemOpen
  \bibfield  {author} {\bibinfo {author} {\bibfnamefont {R.}~\bibnamefont
  {P\"{o}ttgen}}, \bibinfo {author} {\bibfnamefont {H.}~\bibnamefont
  {Borrmann}}, \ and\ \bibinfo {author} {\bibfnamefont {R.~K.}\ \bibnamefont
  {Kremer}},\ }\bibfield  {title} {\enquote {\bibinfo {title} {Ferromagnetic
  ordering in \ce{CeAuGe}},}\ }\href@noop {} {\bibfield  {journal} {\bibinfo
  {journal} {J. Magn. Magn. Mater.}\ }\textbf {\bibinfo {volume} {152}},\
  \bibinfo {pages} {196} (\bibinfo {year} {1996}{\natexlab{b}})}\BibitemShut
  {NoStop}%
\bibitem [{\citenamefont {Gibson}\ \emph {et~al.}(2000)\citenamefont {Gibson},
  \citenamefont {P\"{o}ttgen},\ and\ \citenamefont
  {Kremer}}]{gibson2000magnetic}%
  \BibitemOpen
  \bibfield  {author} {\bibinfo {author} {\bibfnamefont {B.~J.}\ \bibnamefont
  {Gibson}}, \bibinfo {author} {\bibfnamefont {R.}~\bibnamefont {P\"{o}ttgen}},
  \ and\ \bibinfo {author} {\bibfnamefont {R.~K.}\ \bibnamefont {Kremer}},\
  }\bibfield  {title} {\enquote {\bibinfo {title} {Magnetic structure
  determination of \ce{CeAuGe} and \ce{CeAgGe}},}\ }\href@noop {} {\bibfield
  {journal} {\bibinfo  {journal} {Physica B: Condensed Matter}\ }\textbf
  {\bibinfo {volume} {276}},\ \bibinfo {pages} {734} (\bibinfo {year}
  {2000})}\BibitemShut {NoStop}%
\bibitem [{\citenamefont {Gibson}\ \emph {et~al.}(1996)\citenamefont {Gibson},
  \citenamefont {Schnelle}, \citenamefont {P\'{o}ttgen}, \citenamefont
  {Bartkowski},\ and\ \citenamefont {Kremer}}]{gibson1996susceptibility}%
  \BibitemOpen
  \bibfield  {author} {\bibinfo {author} {\bibfnamefont {B.~J.}\ \bibnamefont
  {Gibson}}, \bibinfo {author} {\bibfnamefont {W.}~\bibnamefont {Schnelle}},
  \bibinfo {author} {\bibfnamefont {R.}~\bibnamefont {P\'{o}ttgen}}, \bibinfo
  {author} {\bibfnamefont {K.}~\bibnamefont {Bartkowski}}, \ and\ \bibinfo
  {author} {\bibfnamefont {R.~K.}\ \bibnamefont {Kremer}},\ }\bibfield  {title}
  {\enquote {\bibinfo {title} {Susceptibility, specific heat, and transport
  properties of \ce{CeAuGe} and \ce{GdAuGe}},}\ }\href@noop {} {\bibfield
  {journal} {\bibinfo  {journal} {Czechoslovak Journal of Physics}\ }\textbf
  {\bibinfo {volume} {46}},\ \bibinfo {pages} {2573} (\bibinfo {year}
  {1996})}\BibitemShut {NoStop}%
\bibitem [{\citenamefont {Kaczorowski}\ and\ \citenamefont
  {Szytu{\l}a}(2014)}]{kaczorowski2014magnetic}%
  \BibitemOpen
  \bibfield  {author} {\bibinfo {author} {\bibfnamefont {D.}~\bibnamefont
  {Kaczorowski}}\ and\ \bibinfo {author} {\bibfnamefont {A.}~\bibnamefont
  {Szytu{\l}a}},\ }\bibfield  {title} {\enquote {\bibinfo {title} {Magnetic,
  thermal and electrical transport properties of {T}m{A}u{G}e},}\ }\href@noop
  {} {\bibfield  {journal} {\bibinfo  {journal} {J. Alloys Compd.}\ }\textbf
  {\bibinfo {volume} {614}},\ \bibinfo {pages} {186} (\bibinfo {year}
  {2014})}\BibitemShut {NoStop}%
\bibitem [{\citenamefont {Gibson}(1998)}]{gibson1998investigation}%
  \BibitemOpen
  \bibfield  {author} {\bibinfo {author} {\bibfnamefont {B.~J.}\ \bibnamefont
  {Gibson}},\ }\emph {\bibinfo {title} {Investigation of the physical
  properties of the ternary intermetallic rare-earth compounds, {RETG}e ({RE} =
  {S}c, {Y}, {L}a$-${L}u; {T} = {A}g, {A}u)}},\ \href@noop {} {Ph.D. thesis},\
  \bibinfo  {school} {Loughborough University} (\bibinfo {year}
  {1998})\BibitemShut {NoStop}%
\bibitem [{\citenamefont {Penc}\ \emph {et~al.}(1999)\citenamefont {Penc},
  \citenamefont {Baran}, \citenamefont {\'{S}laski},\ and\ \citenamefont
  {Szytu{\l}a}}]{penc1999magnetic}%
  \BibitemOpen
  \bibfield  {author} {\bibinfo {author} {\bibfnamefont {B.}~\bibnamefont
  {Penc}}, \bibinfo {author} {\bibfnamefont {S.}~\bibnamefont {Baran}},
  \bibinfo {author} {\bibfnamefont {M.}~\bibnamefont {\'{S}laski}}, \ and\
  \bibinfo {author} {\bibfnamefont {A.}~\bibnamefont {Szytu{\l}a}},\ }\bibfield
   {title} {\enquote {\bibinfo {title} {Magnetic properties of
  $\textit{R}${A}u{G}e compounds ($\textit{R}$ = {P}r, {N}d, {T}b$-${E}r)},}\
  }\href@noop {} {\bibfield  {journal} {\bibinfo  {journal} {J. Alloys Compd.}\
  }\textbf {\bibinfo {volume} {282}},\ \bibinfo {pages} {L6} (\bibinfo {year}
  {1999})}\BibitemShut {NoStop}%
\bibitem [{\citenamefont {Baran}\ \emph {et~al.}(2000)\citenamefont {Baran},
  \citenamefont {Hofmann}, \citenamefont {Penc}, \citenamefont {\'{S}laski},
  \citenamefont {Szytu{\l}a},\ and\ \citenamefont
  {Zygmunt}}]{baran2000magnetic}%
  \BibitemOpen
  \bibfield  {author} {\bibinfo {author} {\bibfnamefont {S.}~\bibnamefont
  {Baran}}, \bibinfo {author} {\bibfnamefont {M.}~\bibnamefont {Hofmann}},
  \bibinfo {author} {\bibfnamefont {B.}~\bibnamefont {Penc}}, \bibinfo {author}
  {\bibfnamefont {M.}~\bibnamefont {\'{S}laski}}, \bibinfo {author}
  {\bibfnamefont {A.}~\bibnamefont {Szytu{\l}a}}, \ and\ \bibinfo {author}
  {\bibfnamefont {A.}~\bibnamefont {Zygmunt}},\ }\bibfield  {title} {\enquote
  {\bibinfo {title} {Magnetic structures of $\textit{R}${A}u{G}e ($\textit{R}$
  = {P}r, {N}d, {T}b, {H}o, {E}r) compounds},}\ }\href@noop {} {\bibfield
  {journal} {\bibinfo  {journal} {Physica B}\ }\textbf {\bibinfo {volume}
  {276}},\ \bibinfo {pages} {656} (\bibinfo {year} {2000})}\BibitemShut
  {NoStop}%
\bibitem [{\citenamefont {Baran}\ \emph {et~al.}(2001)\citenamefont {Baran},
  \citenamefont {Hofmann}, \citenamefont {Lampert}, \citenamefont
  {St\"{u}sser}, \citenamefont {Szytu{\l}a}, \citenamefont {T\"{o}bbens},
  \citenamefont {Smeibidl},\ and\ \citenamefont {Kausche}}]{baran2001neutron}%
  \BibitemOpen
  \bibfield  {author} {\bibinfo {author} {\bibfnamefont {S.}~\bibnamefont
  {Baran}}, \bibinfo {author} {\bibfnamefont {M.}~\bibnamefont {Hofmann}},
  \bibinfo {author} {\bibfnamefont {G.}~\bibnamefont {Lampert}}, \bibinfo
  {author} {\bibfnamefont {N.}~\bibnamefont {St\"{u}sser}}, \bibinfo {author}
  {\bibfnamefont {A.}~\bibnamefont {Szytu{\l}a}}, \bibinfo {author}
  {\bibfnamefont {D.}~\bibnamefont {T\"{o}bbens}}, \bibinfo {author}
  {\bibfnamefont {P.}~\bibnamefont {Smeibidl}}, \ and\ \bibinfo {author}
  {\bibfnamefont {S.}~\bibnamefont {Kausche}},\ }\bibfield  {title} {\enquote
  {\bibinfo {title} {Neutron diffraction studies of the magnetic structures of
  {H}o{A}u{G}e and {E}r{A}u{G}e},}\ }\href@noop {} {\bibfield  {journal}
  {\bibinfo  {journal} {J. Magn. Magn. Mater.}\ }\textbf {\bibinfo {volume}
  {236}},\ \bibinfo {pages} {293} (\bibinfo {year} {2001})}\BibitemShut
  {NoStop}%
\bibitem [{\citenamefont {Gibson}\ \emph {et~al.}(2001)\citenamefont {Gibson},
  \citenamefont {P\"{o}ttgen}, \citenamefont {Schnelle}, \citenamefont
  {Ouladdiaf},\ and\ \citenamefont {Kremer}}]{gibson2001crystal}%
  \BibitemOpen
  \bibfield  {author} {\bibinfo {author} {\bibfnamefont {B.~J.}\ \bibnamefont
  {Gibson}}, \bibinfo {author} {\bibfnamefont {R.}~\bibnamefont {P\"{o}ttgen}},
  \bibinfo {author} {\bibfnamefont {W.}~\bibnamefont {Schnelle}}, \bibinfo
  {author} {\bibfnamefont {B.}~\bibnamefont {Ouladdiaf}}, \ and\ \bibinfo
  {author} {\bibfnamefont {R.~K.}\ \bibnamefont {Kremer}},\ }\bibfield  {title}
  {\enquote {\bibinfo {title} {Crystal and magnetic structure of
  antiferromagnetic {H}o{A}u{G}e},}\ }\href@noop {} {\bibfield  {journal}
  {\bibinfo  {journal} {J. Phys.: Condens. Matter}\ }\textbf {\bibinfo {volume}
  {13}},\ \bibinfo {pages} {2593} (\bibinfo {year} {2001})}\BibitemShut
  {NoStop}%
\bibitem [{\citenamefont {Bashir}\ \emph {et~al.}(2014)\citenamefont {Bashir},
  \citenamefont {Tchoula~Tchokont\'{e}}, \citenamefont {Snyman}, \citenamefont
  {Sondezi},\ and\ \citenamefont {Strydom}}]{bashir2014antiferromagnetic}%
  \BibitemOpen
  \bibfield  {author} {\bibinfo {author} {\bibfnamefont {A.~K.~H.}\
  \bibnamefont {Bashir}}, \bibinfo {author} {\bibfnamefont {M.~B.}\
  \bibnamefont {Tchoula~Tchokont\'{e}}}, \bibinfo {author} {\bibfnamefont
  {J.~L.}\ \bibnamefont {Snyman}}, \bibinfo {author} {\bibfnamefont {B.~M.}\
  \bibnamefont {Sondezi}}, \ and\ \bibinfo {author} {\bibfnamefont {A.~M.}\
  \bibnamefont {Strydom}},\ }\bibfield  {title} {\enquote {\bibinfo {title}
  {Antiferromagnetic ordering in \ce{NdAuGe} compound},}\ }\href@noop {}
  {\bibfield  {journal} {\bibinfo  {journal} {J. Appl. Phys.}\ }\textbf
  {\bibinfo {volume} {115}},\ \bibinfo {pages} {17E134} (\bibinfo {year}
  {2014})}\BibitemShut {NoStop}%
\bibitem [{\citenamefont {Schnelle}\ \emph {et~al.}(1997)\citenamefont
  {Schnelle}, \citenamefont {P\"{o}ttgen}, \citenamefont {Kremer},
  \citenamefont {Gmelin},\ and\ \citenamefont {Jepsen}}]{schnelle1997crystal}%
  \BibitemOpen
  \bibfield  {author} {\bibinfo {author} {\bibfnamefont {W.}~\bibnamefont
  {Schnelle}}, \bibinfo {author} {\bibfnamefont {R.}~\bibnamefont
  {P\"{o}ttgen}}, \bibinfo {author} {\bibfnamefont {R.~K.}\ \bibnamefont
  {Kremer}}, \bibinfo {author} {\bibfnamefont {E.}~\bibnamefont {Gmelin}}, \
  and\ \bibinfo {author} {\bibfnamefont {O.}~\bibnamefont {Jepsen}},\
  }\bibfield  {title} {\enquote {\bibinfo {title} {The crystal structure,
  magnetic susceptibility, electrical resistivity, specific heat, and
  electronic band structure of $\textit{R}${A}u{G}e ({R} = {S}c, {Y}, {L}a,
  {L}u)},}\ }\href@noop {} {\bibfield  {journal} {\bibinfo  {journal} {J.
  Phys.: Condens. Matter}\ }\textbf {\bibinfo {volume} {9}},\ \bibinfo {pages}
  {1435} (\bibinfo {year} {1997})}\BibitemShut {NoStop}%
\bibitem [{\citenamefont {Wang}\ \emph {et~al.}(2020)\citenamefont {Wang},
  \citenamefont {Koster}, \citenamefont {Ochs}, \citenamefont {Scudder},
  \citenamefont {Heremans}, \citenamefont {Windl},\ and\ \citenamefont
  {Goldberger}}]{wang2020chemical}%
  \BibitemOpen
  \bibfield  {author} {\bibinfo {author} {\bibfnamefont {Y.}~\bibnamefont
  {Wang}}, \bibinfo {author} {\bibfnamefont {K.~G.}\ \bibnamefont {Koster}},
  \bibinfo {author} {\bibfnamefont {A.~M.}\ \bibnamefont {Ochs}}, \bibinfo
  {author} {\bibfnamefont {M.~R.}\ \bibnamefont {Scudder}}, \bibinfo {author}
  {\bibfnamefont {J.~P.}\ \bibnamefont {Heremans}}, \bibinfo {author}
  {\bibfnamefont {W.}~\bibnamefont {Windl}}, \ and\ \bibinfo {author}
  {\bibfnamefont {J.~E.}\ \bibnamefont {Goldberger}},\ }\bibfield  {title}
  {\enquote {\bibinfo {title} {The chemical design principles for
  axis-dependent conduction polarity},}\ }\href@noop {} {\bibfield  {journal}
  {\bibinfo  {journal} {J. Am. Chem. Soc.}\ }\textbf {\bibinfo {volume}
  {142}},\ \bibinfo {pages} {2812--2822} (\bibinfo {year} {2020})}\BibitemShut
  {NoStop}%
\bibitem [{\citenamefont {Prokofiev}\ and\ \citenamefont
  {Paschen}(2012)}]{prokofiev2012crystal}%
  \BibitemOpen
  \bibfield  {author} {\bibinfo {author} {\bibfnamefont {A.}~\bibnamefont
  {Prokofiev}}\ and\ \bibinfo {author} {\bibfnamefont {S.}~\bibnamefont
  {Paschen}},\ }\bibfield  {title} {\enquote {\bibinfo {title} {Crystal growth
  and stoichiometry of strongly correlated intermetallic cerium compounds},}\
  }in\ \href@noop {} {\emph {\bibinfo {booktitle} {Modern Aspects of Bulk
  Crystal and Thin Film Preparation}}}\ (\bibinfo  {publisher} {IntechOpen},\
  \bibinfo {year} {2012})\BibitemShut {NoStop}%
\bibitem [{\citenamefont {Du}\ \emph {et~al.}(2019)\citenamefont {Du},
  \citenamefont {Lim}, \citenamefont {Zhang}, \citenamefont {Strohbeen},
  \citenamefont {Shourov}, \citenamefont {Rodolakis}, \citenamefont
  {McChesney}, \citenamefont {Voyles}, \citenamefont {Fredrickson},\ and\
  \citenamefont {Kawasaki}}]{du2019high}%
  \BibitemOpen
  \bibfield  {author} {\bibinfo {author} {\bibfnamefont {D.}~\bibnamefont
  {Du}}, \bibinfo {author} {\bibfnamefont {A.}~\bibnamefont {Lim}}, \bibinfo
  {author} {\bibfnamefont {C.}~\bibnamefont {Zhang}}, \bibinfo {author}
  {\bibfnamefont {P.~J.}\ \bibnamefont {Strohbeen}}, \bibinfo {author}
  {\bibfnamefont {E.~H.}\ \bibnamefont {Shourov}}, \bibinfo {author}
  {\bibfnamefont {F.}~\bibnamefont {Rodolakis}}, \bibinfo {author}
  {\bibfnamefont {J.~L.}\ \bibnamefont {McChesney}}, \bibinfo {author}
  {\bibfnamefont {P.}~\bibnamefont {Voyles}}, \bibinfo {author} {\bibfnamefont
  {D.~C.}\ \bibnamefont {Fredrickson}}, \ and\ \bibinfo {author} {\bibfnamefont
  {J.~K.}\ \bibnamefont {Kawasaki}},\ }\bibfield  {title} {\enquote {\bibinfo
  {title} {High electrical conductivity in the epitaxial polar metals
  \ce{LaAuGe} and \ce{LaPtSb}},}\ }\href@noop {} {\bibfield  {journal}
  {\bibinfo  {journal} {APL Materials}\ }\textbf {\bibinfo {volume} {7}},\
  \bibinfo {pages} {121107} (\bibinfo {year} {2019})}\BibitemShut {NoStop}%
\bibitem [{\citenamefont {Stevens}(1952)}]{Stevens}%
  \BibitemOpen
  \bibfield  {author} {\bibinfo {author} {\bibfnamefont {K.~W.~H.}\
  \bibnamefont {Stevens}},\ }\bibfield  {title} {\enquote {\bibinfo {title}
  {Matrix elements and operator equivalents connected with the magnetic
  properties of rare earth ions},}\ }\href@noop {} {\bibfield  {journal}
  {\bibinfo  {journal} {Proc. Phys. Soc. Sect. A}\ }\textbf {\bibinfo {volume}
  {65}},\ \bibinfo {pages} {209} (\bibinfo {year} {1952})}\BibitemShut
  {NoStop}%
\bibitem [{\citenamefont {CrysAlisPRO}(2014)}]{crysalispro2014agilent}%
  \BibitemOpen
  \bibfield  {author} {\bibinfo {author} {\bibfnamefont {Oxford~Diffraction}\
  \bibnamefont {CrysAlisPRO}},\ }\bibfield  {title} {\enquote {\bibinfo {title}
  {Agilent {T}echnologies {UK} {L}td},}\ }\href@noop {} {\bibfield  {journal}
  {\bibinfo  {journal} {Yarnton, England}\ }\textbf {\bibinfo {volume} {1}}
  (\bibinfo {year} {2014})}\BibitemShut {NoStop}%
\bibitem [{\citenamefont {Palatinus}\ and\ \citenamefont
  {Chapuis}(2007)}]{palatinus2007superflip}%
  \BibitemOpen
  \bibfield  {author} {\bibinfo {author} {\bibfnamefont {L.}~\bibnamefont
  {Palatinus}}\ and\ \bibinfo {author} {\bibfnamefont {G.}~\bibnamefont
  {Chapuis}},\ }\bibfield  {title} {\enquote {\bibinfo {title} {{SUPERFLIP}--a
  computer program for the solution of crystal structures by charge flipping in
  arbitrary dimensions},}\ }\href@noop {} {\bibfield  {journal} {\bibinfo
  {journal} {J. Appl. Cryst.}\ }\textbf {\bibinfo {volume} {40}},\ \bibinfo
  {pages} {786--790} (\bibinfo {year} {2007})}\BibitemShut {NoStop}%
\bibitem [{\citenamefont {Pet{\v{r}}{\'\i}{\v{c}}ek}\ \emph
  {et~al.}(2014)\citenamefont {Pet{\v{r}}{\'\i}{\v{c}}ek}, \citenamefont
  {Du{\v{s}}ek},\ and\ \citenamefont
  {Palatinus}}]{petvrivcek2014crystallographic}%
  \BibitemOpen
  \bibfield  {author} {\bibinfo {author} {\bibfnamefont {V{\'a}clav}\
  \bibnamefont {Pet{\v{r}}{\'\i}{\v{c}}ek}}, \bibinfo {author} {\bibfnamefont
  {Michal}\ \bibnamefont {Du{\v{s}}ek}}, \ and\ \bibinfo {author}
  {\bibfnamefont {Luk{\'a}{\v{s}}}\ \bibnamefont {Palatinus}},\ }\bibfield
  {title} {\enquote {\bibinfo {title} {Crystallographic computing system
  {JANA}2006: general features},}\ }\href@noop {} {\bibfield  {journal}
  {\bibinfo  {journal} {Z. Kristallogr. Cryst. Mater.}\ }\textbf {\bibinfo
  {volume} {229}},\ \bibinfo {pages} {345--352} (\bibinfo {year}
  {2014})}\BibitemShut {NoStop}%
\bibitem [{\citenamefont {Merlo}\ \emph {et~al.}(1998)\citenamefont {Merlo},
  \citenamefont {Pani}, \citenamefont {Canepa},\ and\ \citenamefont
  {Fornasini}}]{merlo1998phases}%
  \BibitemOpen
  \bibfield  {author} {\bibinfo {author} {\bibfnamefont {F.}~\bibnamefont
  {Merlo}}, \bibinfo {author} {\bibfnamefont {M.}~\bibnamefont {Pani}},
  \bibinfo {author} {\bibfnamefont {F.}~\bibnamefont {Canepa}}, \ and\ \bibinfo
  {author} {\bibfnamefont {M.~L.}\ \bibnamefont {Fornasini}},\ }\bibfield
  {title} {\enquote {\bibinfo {title} {Phases around the 1:1:1 composition in
  the {Y}b-{A}u-{G}e and {C}a-{A}u-{G}e systems},}\ }\href@noop {} {\bibfield
  {journal} {\bibinfo  {journal} {J. Alloys Compd.}\ }\textbf {\bibinfo
  {volume} {264}},\ \bibinfo {pages} {82--88} (\bibinfo {year}
  {1998})}\BibitemShut {NoStop}%
\bibitem [{\citenamefont {Kajimoto}\ \emph {et~al.}(2011)\citenamefont
  {Kajimoto}, \citenamefont {Nakamura}, \citenamefont {Inamura}, \citenamefont
  {Mizuno}, \citenamefont {Nakajima}, \citenamefont {Ohira-Kawamura},
  \citenamefont {Yokoo}, \citenamefont {Nakatani}, \citenamefont {Maruyama},
  \citenamefont {Soyama}, \citenamefont {Shibata}, \citenamefont {Suzuya},
  \citenamefont {Sato}, \citenamefont {Aizawa}, \citenamefont {Arai},
  \citenamefont {Wakimoto}, \citenamefont {Ishikado}, \citenamefont {Shamoto},
  \citenamefont {Fujita}, \citenamefont {Hiraka}, \citenamefont {Ohoyama},
  \citenamefont {Yamada},\ and\ \citenamefont {Lee}}]{4seasons}%
  \BibitemOpen
  \bibfield  {author} {\bibinfo {author} {\bibfnamefont {R.}~\bibnamefont
  {Kajimoto}}, \bibinfo {author} {\bibfnamefont {M.}~\bibnamefont {Nakamura}},
  \bibinfo {author} {\bibfnamefont {Y.}~\bibnamefont {Inamura}}, \bibinfo
  {author} {\bibfnamefont {F.}~\bibnamefont {Mizuno}}, \bibinfo {author}
  {\bibfnamefont {K.}~\bibnamefont {Nakajima}}, \bibinfo {author}
  {\bibfnamefont {S.}~\bibnamefont {Ohira-Kawamura}}, \bibinfo {author}
  {\bibfnamefont {T.}~\bibnamefont {Yokoo}}, \bibinfo {author} {\bibfnamefont
  {T.}~\bibnamefont {Nakatani}}, \bibinfo {author} {\bibfnamefont
  {R.}~\bibnamefont {Maruyama}}, \bibinfo {author} {\bibfnamefont
  {K.}~\bibnamefont {Soyama}}, \bibinfo {author} {\bibfnamefont
  {K.}~\bibnamefont {Shibata}}, \bibinfo {author} {\bibfnamefont
  {K.}~\bibnamefont {Suzuya}}, \bibinfo {author} {\bibfnamefont
  {S.}~\bibnamefont {Sato}}, \bibinfo {author} {\bibfnamefont {K.}~\bibnamefont
  {Aizawa}}, \bibinfo {author} {\bibfnamefont {M.}~\bibnamefont {Arai}},
  \bibinfo {author} {\bibfnamefont {S.}~\bibnamefont {Wakimoto}}, \bibinfo
  {author} {\bibfnamefont {M.}~\bibnamefont {Ishikado}}, \bibinfo {author}
  {\bibfnamefont {S.-i.}\ \bibnamefont {Shamoto}}, \bibinfo {author}
  {\bibfnamefont {M.}~\bibnamefont {Fujita}}, \bibinfo {author} {\bibfnamefont
  {H.}~\bibnamefont {Hiraka}}, \bibinfo {author} {\bibfnamefont
  {K.}~\bibnamefont {Ohoyama}}, \bibinfo {author} {\bibfnamefont
  {K.}~\bibnamefont {Yamada}}, \ and\ \bibinfo {author} {\bibfnamefont {C.-H.}\
  \bibnamefont {Lee}},\ }\bibfield  {title} {\enquote {\bibinfo {title} {The
  {F}ermi chopper spectrometer 4{SEASONS} at {J-PARC}},}\ }\href@noop {}
  {\bibfield  {journal} {\bibinfo  {journal} {J. Phys. Soc. Jpn.}\ }\textbf
  {\bibinfo {volume} {80}},\ \bibinfo {pages} {SB025} (\bibinfo {year}
  {2011})}\BibitemShut {NoStop}%
\bibitem [{\citenamefont {Inamura}\ \emph {et~al.}(2013)\citenamefont
  {Inamura}, \citenamefont {Nakatani}, \citenamefont {Suzuki},\ and\
  \citenamefont {Otomo}}]{UTSUSEMI}%
  \BibitemOpen
  \bibfield  {author} {\bibinfo {author} {\bibfnamefont {Y.}~\bibnamefont
  {Inamura}}, \bibinfo {author} {\bibfnamefont {T.}~\bibnamefont {Nakatani}},
  \bibinfo {author} {\bibfnamefont {J.}~\bibnamefont {Suzuki}}, \ and\ \bibinfo
  {author} {\bibfnamefont {T.}~\bibnamefont {Otomo}},\ }\bibfield  {title}
  {\enquote {\bibinfo {title} {Development status of software “{U}tsusemi”
  for chopper spectrometers at {MLF}, {J-PARC}},}\ }\href@noop {} {\bibfield
  {journal} {\bibinfo  {journal} {J. Phys. Soc. Jpn.}\ }\textbf {\bibinfo
  {volume} {82}},\ \bibinfo {pages} {SA031} (\bibinfo {year}
  {2013})}\BibitemShut {NoStop}%
\bibitem [{\citenamefont {Shannon}(1976)}]{shannon1976revised}%
  \BibitemOpen
  \bibfield  {author} {\bibinfo {author} {\bibfnamefont {R.~D.}\ \bibnamefont
  {Shannon}},\ }\bibfield  {title} {\enquote {\bibinfo {title} {Revised
  effective ionic radii and systematic studies of interatomic distances in
  halides and chalcogenides},}\ }\href@noop {} {\bibfield  {journal} {\bibinfo
  {journal} {Acta Crystallogr. A}\ }\textbf {\bibinfo {volume} {32}},\ \bibinfo
  {pages} {751--767} (\bibinfo {year} {1976})}\BibitemShut {NoStop}%
\bibitem [{\citenamefont {Hoffmann}\ and\ \citenamefont
  {P\"{o}ttgen}(2001)}]{hoffmann2001alb2}%
  \BibitemOpen
  \bibfield  {author} {\bibinfo {author} {\bibfnamefont {R.-D.}\ \bibnamefont
  {Hoffmann}}\ and\ \bibinfo {author} {\bibfnamefont {Rainer}\ \bibnamefont
  {P\"{o}ttgen}},\ }\bibfield  {title} {\enquote {\bibinfo {title}
  {\ce{AlB2}-related intermetallic compounds-a comprehensive view based on
  group-subgroup relations},}\ }\href@noop {} {\bibfield  {journal} {\bibinfo
  {journal} {Z. Kristallogr. Cryst. Mater.}\ }\textbf {\bibinfo {volume}
  {216}},\ \bibinfo {pages} {127} (\bibinfo {year} {2001})}\BibitemShut
  {NoStop}%
\bibitem [{\citenamefont {Wang}(1971)}]{wang1971crystal}%
  \BibitemOpen
  \bibfield  {author} {\bibinfo {author} {\bibfnamefont {Y.-L.}\ \bibnamefont
  {Wang}},\ }\bibfield  {title} {\enquote {\bibinfo {title} {Crystal-field
  effects of paramagnetic {C}urie temperature},}\ }\href@noop {} {\bibfield
  {journal} {\bibinfo  {journal} {Phys. Lett. A}\ }\textbf {\bibinfo {volume}
  {35}},\ \bibinfo {pages} {383--384} (\bibinfo {year} {1971})}\BibitemShut
  {NoStop}%
\bibitem [{\citenamefont {Boutron}(1973)}]{boutron1973exact}%
  \BibitemOpen
  \bibfield  {author} {\bibinfo {author} {\bibfnamefont {P.}~\bibnamefont
  {Boutron}},\ }\bibfield  {title} {\enquote {\bibinfo {title} {Exact
  calculation of the paramagnetic susceptibility of a single crystal with
  arbitrary crystal field and exchange interactions},}\ }\href@noop {}
  {\bibfield  {journal} {\bibinfo  {journal} {Phys. Rev. B}\ }\textbf {\bibinfo
  {volume} {7}},\ \bibinfo {pages} {3226} (\bibinfo {year} {1973})}\BibitemShut
  {NoStop}%
\bibitem [{\citenamefont {Van~Hieu}\ \emph {et~al.}(2007)\citenamefont
  {Van~Hieu}, \citenamefont {Takeuchi}, \citenamefont {Shishido}, \citenamefont
  {Tonohiro}, \citenamefont {Yamada}, \citenamefont {Nakashima}, \citenamefont
  {Sugiyama}, \citenamefont {Settai}, \citenamefont {Matsuda}, \citenamefont
  {Haga} \emph {et~al.}}]{van2007magnetic}%
  \BibitemOpen
  \bibfield  {author} {\bibinfo {author} {\bibfnamefont {N.}~\bibnamefont
  {Van~Hieu}}, \bibinfo {author} {\bibfnamefont {T.}~\bibnamefont {Takeuchi}},
  \bibinfo {author} {\bibfnamefont {H.}~\bibnamefont {Shishido}}, \bibinfo
  {author} {\bibfnamefont {C.}~\bibnamefont {Tonohiro}}, \bibinfo {author}
  {\bibfnamefont {T.}~\bibnamefont {Yamada}}, \bibinfo {author} {\bibfnamefont
  {H.}~\bibnamefont {Nakashima}}, \bibinfo {author} {\bibfnamefont
  {K.}~\bibnamefont {Sugiyama}}, \bibinfo {author} {\bibfnamefont
  {R.}~\bibnamefont {Settai}}, \bibinfo {author} {\bibfnamefont {T.~D.}\
  \bibnamefont {Matsuda}}, \bibinfo {author} {\bibfnamefont {Y.}~\bibnamefont
  {Haga}},  \emph {et~al.},\ }\bibfield  {title} {\enquote {\bibinfo {title}
  {Magnetic properties and crystalline electric field scheme in
  $\textit{R}$\ce{RhIn5} ($\textit{R}$: Rare earth)},}\ }\href@noop {}
  {\bibfield  {journal} {\bibinfo  {journal} {J. Phys. Soc. Jpn.}\ }\textbf
  {\bibinfo {volume} {76}},\ \bibinfo {pages} {064702} (\bibinfo {year}
  {2007})}\BibitemShut {NoStop}%
\bibitem [{\citenamefont {Li}\ \emph {et~al.}(2020)\citenamefont {Li},
  \citenamefont {Bachus}, \citenamefont {Deng}, \citenamefont {Schmidt},
  \citenamefont {Thoma}, \citenamefont {Hutanu}, \citenamefont {Tokiwa},
  \citenamefont {Tsirlin},\ and\ \citenamefont {Gegenwart}}]{li2020partial}%
  \BibitemOpen
  \bibfield  {author} {\bibinfo {author} {\bibfnamefont {Y.}~\bibnamefont
  {Li}}, \bibinfo {author} {\bibfnamefont {S.}~\bibnamefont {Bachus}}, \bibinfo
  {author} {\bibfnamefont {H.}~\bibnamefont {Deng}}, \bibinfo {author}
  {\bibfnamefont {W.}~\bibnamefont {Schmidt}}, \bibinfo {author} {\bibfnamefont
  {H.}~\bibnamefont {Thoma}}, \bibinfo {author} {\bibfnamefont
  {V.}~\bibnamefont {Hutanu}}, \bibinfo {author} {\bibfnamefont
  {Y.}~\bibnamefont {Tokiwa}}, \bibinfo {author} {\bibfnamefont {A.~A.}\
  \bibnamefont {Tsirlin}}, \ and\ \bibinfo {author} {\bibfnamefont
  {P.}~\bibnamefont {Gegenwart}},\ }\bibfield  {title} {\enquote {\bibinfo
  {title} {Partial up-up-down order with the continuously distributed order
  parameter in the triangular antiferromagnet \ce{TmMgGaO4}},}\ }\href@noop {}
  {\bibfield  {journal} {\bibinfo  {journal} {Phys. Rev. X}\ }\textbf {\bibinfo
  {volume} {10}},\ \bibinfo {pages} {011007} (\bibinfo {year}
  {2020})}\BibitemShut {NoStop}%
\bibitem [{\citenamefont {Shen}\ \emph {et~al.}(2019)\citenamefont {Shen},
  \citenamefont {Liu}, \citenamefont {Qin}, \citenamefont {Shen}, \citenamefont
  {Li}, \citenamefont {Bewley}, \citenamefont {Schneidewind}, \citenamefont
  {Chen},\ and\ \citenamefont {Zhao}}]{shen2019intertwined}%
  \BibitemOpen
  \bibfield  {author} {\bibinfo {author} {\bibfnamefont {Y.}~\bibnamefont
  {Shen}}, \bibinfo {author} {\bibfnamefont {C.}~\bibnamefont {Liu}}, \bibinfo
  {author} {\bibfnamefont {Y.}~\bibnamefont {Qin}}, \bibinfo {author}
  {\bibfnamefont {S.}~\bibnamefont {Shen}}, \bibinfo {author} {\bibfnamefont
  {Y.-D.}\ \bibnamefont {Li}}, \bibinfo {author} {\bibfnamefont
  {R.}~\bibnamefont {Bewley}}, \bibinfo {author} {\bibfnamefont
  {A.}~\bibnamefont {Schneidewind}}, \bibinfo {author} {\bibfnamefont
  {G.}~\bibnamefont {Chen}}, \ and\ \bibinfo {author} {\bibfnamefont
  {J.}~\bibnamefont {Zhao}},\ }\bibfield  {title} {\enquote {\bibinfo {title}
  {Intertwined dipolar and multipolar order in the triangular-lattice magnet
  \ce{TmMgGaO4}},}\ }\href@noop {} {\bibfield  {journal} {\bibinfo  {journal}
  {Nat. Commun.}\ }\textbf {\bibinfo {volume} {10}},\ \bibinfo {pages} {1--7}
  (\bibinfo {year} {2019})}\BibitemShut {NoStop}%
\bibitem [{\citenamefont {Doerr}\ \emph {et~al.}(2005)\citenamefont {Doerr},
  \citenamefont {Rotter},\ and\ \citenamefont
  {Lindbaum}}]{doerr2005magnetostriction}%
  \BibitemOpen
  \bibfield  {author} {\bibinfo {author} {\bibfnamefont {M.}~\bibnamefont
  {Doerr}}, \bibinfo {author} {\bibfnamefont {M.}~\bibnamefont {Rotter}}, \
  and\ \bibinfo {author} {\bibfnamefont {A.}~\bibnamefont {Lindbaum}},\
  }\bibfield  {title} {\enquote {\bibinfo {title} {Magnetostriction in
  rare-earth based antiferromagnets},}\ }\href@noop {} {\bibfield  {journal}
  {\bibinfo  {journal} {Adv. Phys.}\ }\textbf {\bibinfo {volume} {54}},\
  \bibinfo {pages} {1} (\bibinfo {year} {2005})}\BibitemShut {NoStop}%
\bibitem [{\citenamefont {Grover}(1965)}]{grover1965dynamical}%
  \BibitemOpen
  \bibfield  {author} {\bibinfo {author} {\bibfnamefont {B.}~\bibnamefont
  {Grover}},\ }\bibfield  {title} {\enquote {\bibinfo {title} {Dynamical
  properties of induced-moment systems},}\ }\href@noop {} {\bibfield  {journal}
  {\bibinfo  {journal} {Phys. Rev.}\ }\textbf {\bibinfo {volume} {140}},\
  \bibinfo {pages} {A1944} (\bibinfo {year} {1965})}\BibitemShut {NoStop}%
\bibitem [{\citenamefont {Goremychkin}\ \emph {et~al.}(2008)\citenamefont
  {Goremychkin}, \citenamefont {Osborn}, \citenamefont {Rainford},
  \citenamefont {Macaluso}, \citenamefont {Adroja},\ and\ \citenamefont
  {Koza}}]{goremychkin2008spin}%
  \BibitemOpen
  \bibfield  {author} {\bibinfo {author} {\bibfnamefont {E.~A.}\ \bibnamefont
  {Goremychkin}}, \bibinfo {author} {\bibfnamefont {R.}~\bibnamefont {Osborn}},
  \bibinfo {author} {\bibfnamefont {B.~D.}\ \bibnamefont {Rainford}}, \bibinfo
  {author} {\bibfnamefont {R.~T.}\ \bibnamefont {Macaluso}}, \bibinfo {author}
  {\bibfnamefont {D.~T.}\ \bibnamefont {Adroja}}, \ and\ \bibinfo {author}
  {\bibfnamefont {M.}~\bibnamefont {Koza}},\ }\bibfield  {title} {\enquote
  {\bibinfo {title} {Spin-glass order induced by dynamic frustration},}\
  }\href@noop {} {\bibfield  {journal} {\bibinfo  {journal} {Nat. Phys.}\
  }\textbf {\bibinfo {volume} {4}},\ \bibinfo {pages} {766--770} (\bibinfo
  {year} {2008})}\BibitemShut {NoStop}%
\bibitem [{\citenamefont {Andres}\ \emph {et~al.}(1972)\citenamefont {Andres},
  \citenamefont {Bucher}, \citenamefont {Darack},\ and\ \citenamefont
  {Maita}}]{andres1972induced}%
  \BibitemOpen
  \bibfield  {author} {\bibinfo {author} {\bibfnamefont {K.}~\bibnamefont
  {Andres}}, \bibinfo {author} {\bibfnamefont {E.}~\bibnamefont {Bucher}},
  \bibinfo {author} {\bibfnamefont {S.}~\bibnamefont {Darack}}, \ and\ \bibinfo
  {author} {\bibfnamefont {J.~P.}\ \bibnamefont {Maita}},\ }\bibfield  {title}
  {\enquote {\bibinfo {title} {Induced-moment ferromagnetism in \ce{Pr3Tl}},}\
  }\href@noop {} {\bibfield  {journal} {\bibinfo  {journal} {Phys. Rev. B}\
  }\textbf {\bibinfo {volume} {6}},\ \bibinfo {pages} {2716} (\bibinfo {year}
  {1972})}\BibitemShut {NoStop}%
\bibitem [{\citenamefont {Kitazawa}\ \emph {et~al.}(1998)\citenamefont
  {Kitazawa}, \citenamefont {D\"{o}nni}, \citenamefont {Keller}, \citenamefont
  {Tang}, \citenamefont {Fauth},\ and\ \citenamefont
  {Kido}}]{kitazawa1998magnetic}%
  \BibitemOpen
  \bibfield  {author} {\bibinfo {author} {\bibfnamefont {H.}~\bibnamefont
  {Kitazawa}}, \bibinfo {author} {\bibfnamefont {A.}~\bibnamefont {D\"{o}nni}},
  \bibinfo {author} {\bibfnamefont {L.}~\bibnamefont {Keller}}, \bibinfo
  {author} {\bibfnamefont {J.}~\bibnamefont {Tang}}, \bibinfo {author}
  {\bibfnamefont {F.}~\bibnamefont {Fauth}}, \ and\ \bibinfo {author}
  {\bibfnamefont {G.}~\bibnamefont {Kido}},\ }\bibfield  {title} {\enquote
  {\bibinfo {title} {Magnetic structures of the rare-earth platinum aluminides
  $\textit{R}${P}t{A}l ($\textit{R}$ = {C}e, {P}r, {N}d)},}\ }\href@noop {}
  {\bibfield  {journal} {\bibinfo  {journal} {J. Solid State Chem.}\ }\textbf
  {\bibinfo {volume} {140}},\ \bibinfo {pages} {233--241} (\bibinfo {year}
  {1998})}\BibitemShut {NoStop}%
\bibitem [{\citenamefont {Anand}\ \emph {et~al.}(2014)\citenamefont {Anand},
  \citenamefont {Adroja}, \citenamefont {Bhattacharyya}, \citenamefont
  {Hillier}, \citenamefont {Taylor},\ and\ \citenamefont
  {Strydom}}]{anand2014investigations}%
  \BibitemOpen
  \bibfield  {author} {\bibinfo {author} {\bibfnamefont {V.~K.}\ \bibnamefont
  {Anand}}, \bibinfo {author} {\bibfnamefont {D.~T.}\ \bibnamefont {Adroja}},
  \bibinfo {author} {\bibfnamefont {A.}~\bibnamefont {Bhattacharyya}}, \bibinfo
  {author} {\bibfnamefont {A.~D.}\ \bibnamefont {Hillier}}, \bibinfo {author}
  {\bibfnamefont {J.~W.}\ \bibnamefont {Taylor}}, \ and\ \bibinfo {author}
  {\bibfnamefont {A.~M.}\ \bibnamefont {Strydom}},\ }\bibfield  {title}
  {\enquote {\bibinfo {title} {Investigations of the singlet ground state
  system: \ce{PrIrSi3}},}\ }\href@noop {} {\bibfield  {journal} {\bibinfo
  {journal} {J. Phys.: Condens. Matter}\ }\textbf {\bibinfo {volume} {26}},\
  \bibinfo {pages} {306001} (\bibinfo {year} {2014})}\BibitemShut {NoStop}%
\bibitem [{\citenamefont {De~Gennes}(1962)}]{de1962indirect}%
  \BibitemOpen
  \bibfield  {author} {\bibinfo {author} {\bibfnamefont {P.~G.}\ \bibnamefont
  {De~Gennes}},\ }\bibfield  {title} {\enquote {\bibinfo {title} {Indirect
  interactions between 4f shells in rare earth metals},}\ }\href@noop {}
  {\bibfield  {journal} {\bibinfo  {journal} {J. Phys. Radium}\ }\textbf
  {\bibinfo {volume} {23}},\ \bibinfo {pages} {510} (\bibinfo {year}
  {1962})}\BibitemShut {NoStop}%
\bibitem [{\citenamefont {Liu}\ and\ \citenamefont
  {Guo}(2003)}]{liu2003transition}%
  \BibitemOpen
  \bibfield  {author} {\bibinfo {author} {\bibfnamefont {Z.-S.}\ \bibnamefont
  {Liu}}\ and\ \bibinfo {author} {\bibfnamefont {S.-L.}\ \bibnamefont {Guo}},\
  }\bibfield  {title} {\enquote {\bibinfo {title} {Transition temperatures of
  rare-earth compounds studied with perturbation theory},}\ }\href@noop {}
  {\bibfield  {journal} {\bibinfo  {journal} {Phys. Lett. A}\ }\textbf
  {\bibinfo {volume} {314}},\ \bibinfo {pages} {491--497} (\bibinfo {year}
  {2003})}\BibitemShut {NoStop}%
\bibitem [{\citenamefont {Noakes}\ and\ \citenamefont
  {Shenoy}(1982)}]{noakes1982effect}%
  \BibitemOpen
  \bibfield  {author} {\bibinfo {author} {\bibfnamefont {D.~R.}\ \bibnamefont
  {Noakes}}\ and\ \bibinfo {author} {\bibfnamefont {G.~K.}\ \bibnamefont
  {Shenoy}},\ }\bibfield  {title} {\enquote {\bibinfo {title} {The effect of a
  crystalline electric field on the magnetic transition temperatures of
  rare-earth rhodium borides},}\ }\href@noop {} {\bibfield  {journal} {\bibinfo
   {journal} {Phys. Lett. A}\ }\textbf {\bibinfo {volume} {91}},\ \bibinfo
  {pages} {35--36} (\bibinfo {year} {1982})}\BibitemShut {NoStop}%
\bibitem [{\citenamefont {Luong}\ \emph {et~al.}(2001)\citenamefont {Luong},
  \citenamefont {Franse},\ and\ \citenamefont {Hai}}]{luong2001effect}%
  \BibitemOpen
  \bibfield  {author} {\bibinfo {author} {\bibfnamefont {N.~H.}\ \bibnamefont
  {Luong}}, \bibinfo {author} {\bibfnamefont {J.~J.~M.}\ \bibnamefont
  {Franse}}, \ and\ \bibinfo {author} {\bibfnamefont {N.~H.}\ \bibnamefont
  {Hai}},\ }\bibfield  {title} {\enquote {\bibinfo {title} {Effect of the
  crystalline electric field on the {N}\'{e}el temperatures of
  $\textit{R}${C}u$_2$ compounds},}\ }\href@noop {} {\bibfield  {journal}
  {\bibinfo  {journal} {J. Magn. Magn. Mater.}\ }\textbf {\bibinfo {volume}
  {224}},\ \bibinfo {pages} {30--32} (\bibinfo {year} {2001})}\BibitemShut
  {NoStop}%
\bibitem [{\citenamefont {Lines}(1975)}]{lines1975sensitivity}%
  \BibitemOpen
  \bibfield  {author} {\bibinfo {author} {\bibfnamefont {M.~E.}\ \bibnamefont
  {Lines}},\ }\bibfield  {title} {\enquote {\bibinfo {title} {Sensitivity of
  {C}urie temperature to single-ion anisotropy},}\ }\href@noop {} {\bibfield
  {journal} {\bibinfo  {journal} {Phys. Rev. B}\ }\textbf {\bibinfo {volume}
  {12}},\ \bibinfo {pages} {3766} (\bibinfo {year} {1975})}\BibitemShut
  {NoStop}%
\bibitem [{\citenamefont {Hickox-Young}\ \emph {et~al.}(2022)\citenamefont
  {Hickox-Young}, \citenamefont {Puggioni},\ and\ \citenamefont
  {Rondinelli}}]{hickox2022polar}%
  \BibitemOpen
  \bibfield  {author} {\bibinfo {author} {\bibfnamefont {D.}~\bibnamefont
  {Hickox-Young}}, \bibinfo {author} {\bibfnamefont {D.}~\bibnamefont
  {Puggioni}}, \ and\ \bibinfo {author} {\bibfnamefont {J.~M.}\ \bibnamefont
  {Rondinelli}},\ }\bibfield  {title} {\enquote {\bibinfo {title} {Polar metals
  taxonomy for materials classification and discovery},}\ }\href@noop {}
  {\bibfield  {journal} {\bibinfo  {journal} {arXiv preprint arXiv:2210.05110}\
  } (\bibinfo {year} {2022})}\BibitemShut {NoStop}%
\bibitem [{\citenamefont {Bhowal}\ and\ \citenamefont
  {Spaldin}(2022)}]{bhowal2022polar}%
  \BibitemOpen
  \bibfield  {author} {\bibinfo {author} {\bibfnamefont {S.}~\bibnamefont
  {Bhowal}}\ and\ \bibinfo {author} {\bibfnamefont {N.~A.}\ \bibnamefont
  {Spaldin}},\ }\bibfield  {title} {\enquote {\bibinfo {title} {Polar metals:
  principles and prospects},}\ }\href@noop {} {\bibfield  {journal} {\bibinfo
  {journal} {arXiv preprint arXiv:2210.02993}\ } (\bibinfo {year}
  {2022})}\BibitemShut {NoStop}%
\end{thebibliography}%

\end{document}